\documentclass[preprintnumbers,amsmath,amssymb,prfluids,]{revtex4-2}

\usepackage{graphicx}
\usepackage{soul}
\usepackage{float}
\usepackage{dcolumn}
\usepackage{bm}
\usepackage[latin1]{inputenc}
\usepackage{xcolor}

\begin{document}


\title{Rational constitutive law for the viscous stress tensor in incompressible two-phase flows:
Derivation and tests against a 3D benchmark experiment}

\author{Jacques Magnaudet} \email{jmagnaud@imft.fr}
\author{Hadrien Bruhier}
\author{Samuel Mer}\altaffiliation[Also at ]{PROMES Laboratory (UPR 8521), CNRS - University of Perpignan (UPVD), 66100 Perpignan, France.}
\author{Thomas Bonometti}
\affiliation{Institut de M\'ecanique des Fluides de Toulouse (IMFT), University of Toulouse, CNRS, 31400 Toulouse, France}

\date{\today}

\begin{abstract}
We analyze the representation of viscous stresses in the one-fluid formulation of the two-phase Navier-Stokes equations, the model on which all computational approaches making use of a fixed mesh to discretize the flow field are grounded. Recognizing that the Navier-Stokes-like equations that are actually solved in these approaches imply a spatial filtering, we show by considering a specific two-dimensional flow configuration that the proper representation of the viscous stress tensor requires the introduction of two distinct viscosity coefficients, owing to the different behaviors of shear and normal stresses in control volumes straddling the interface. Making use of classical results of continuum mechanics for anisotropic fluids, we derive the general form of the constitutive law linking the viscous stress tensor of the two-phase medium to the filtered strain-rate tensor, and take advantage of the findings provided by the above two-dimensional configuration to close the determination of the fluid-dependent coefficients involved. Predictions of the resulting anisotropic model are then assessed and compared with those of available \textit{ad hoc} models against original experimental results obtained in a reference flow in which some parts of the interface are dominated by shear while others are mostly controlled by stretching. The selected configuration corresponds to a viscous buoyancy-driven exchange flow in a closed vertical pipe, generated by unstably superimposing two immiscible fluids with a large viscosity contrast and negligible interfacial tension and molecular diffusivity. 
\textcolor{black}{Using different levels of grid refinement}, we show that the anisotropic model is the only one capable of predicting correctly the evolution of the front of the ascending and descending fingers \textcolor{black}{at a reasonable computational cost. \textcolor{black}{We confirm this conclusion on a one-dimensional counter-current Poiseuille flow with a large viscosity contrast, and use this simple configuration to establish that the anisotropic model converges quadratically as the grid is refined, while the familiar \textit{ad hoc} model based on an arithmetic averaging of the fluid viscosities only converges linearly}}.

\end{abstract}

\maketitle

\section{Introduction}
Computational studies of incompressible, viscous, two-phase flows in the presence of surface tension effects have flourished in fluid dynamics research, as they open up endless possibilities for deciphering subtle phenomena and predicting the behavior of complex flows involved in multiple systems of environmental or engineering relevance (see \cite{Prosperetti07,Popinet18} for reviews). 
In most cases, the flow is computed on a fixed grid on which the interfaces are moving freely. Such fixed-grid methods (FGM) have the decisive advantage that the shape of interfaces and possibly their topology can evolve freely over time, so that reconfigurations of the flow structure and, if necessary, break-up and coalescence phenomena, can be considered. FGM originate in the work of Noh and Woodward \cite{Noh76} and Hirt and Nichols \cite{Hirt81} who established the basis of the Volume Of Fluid (VOF) method. Since then, this method has become increasingly popular and has been improved in many respects; see e.g. \cite*{Scardovelli99,Tryggvason11}. Alternative approaches such as the Level Set method \cite*{Sussman94, Sethian96, Sethian03, Gibou18} or the Front Tracking method \cite*{Unverdi92, Tryggvason01} have also enlightened many aspects of dispersed two-phase flows and interfacial fluid dynamics, including phase change phenomena. Also, the Immersed Boundary Method (IBM) initially developed by Peskin \cite{Peskin77} in the context of heart biomechanics is now routinely employed to compute flows around rigid and deformable particles and tackle complex fluid-structure interaction problems \citep{Mittal05, Griffith20, Verzicco23}. \\
\indent Each of these methods has its advantages and shortcomings but they all share certain features. In every FGM, each elementary cell of the computational grid may contain both fluids at a given time. Within such a cell, the two-phase medium appears as a mixture whose physical properties have to be evaluated by averaging according to a certain rule those of the two `pure' fluids. Because of the presence of such `mixed' cells, taking properly into account the whole set of boundary conditions that must be satisfied at an interface separating two viscous fluids is not obvious. 
Formulations relating the geometrical properties of the interface to smooth field variables have been derived for that purpose. In particular, this has led to a continuous formulation of the capillary force, which can then be treated as a usual source term in the Navier-Stokes equation rather than a singular term only defined on the interface \cite{Brackbill92}. \\
\indent
Another difficulty of FGM lies in the evaluation of viscous stresses in the region straddling the interface. In particular, it is necessary to determine how the viscosity depends on those of the two pure fluids according to the local composition of the two-phase medium. Although this issue has long been identified \cite{Kothe98}, it has received only limited attention, possibly because, for several decades, the viscosity ratio between the two fluids was most of the time kept moderate to avoid large gradients and time-step constraints. An exception is the one-dimensional mathematical analysis of the consequences of an inadequate evaluation of heat fluxes and viscous stresses in FGM carried out by Ferziger \cite{Ferziger03}. Some studies performed in two- and three-dimensional configurations also reported consequences of this inadequacy. For instance, the sedimentation velocity of a periodic array of drops moving at low Reynolds number in a less viscous liquid was computed in \cite{Scardovelli99} using a VOF method. Good agreement with theoretical predictions was obtained when the viscosity ratio was of $\mathcal{O}(1)$, but the predicted drop speed was found to under-estimate the theoretical value as the viscosity ratio increased, with a difference already of the order of $10\%$ with a viscosity ratio of $10$. A tougher test was carried out in \cite{Coward97} with the aim of predicting
the growth rate of interfacial disturbances driven by viscosity stratification in a two-layer Couette flow. To obtain numerical predictions consistent with those of linear stability theory, authors had to introduce \textit{ad hoc} modifications in the evaluation of the viscosity in the interfacial region, with distinct weighting rules for evaluating the viscosity involved in the tangential (shear) stress and that involved in the normal stresses. \\
\indent Clearly, the current practice regarding the modeling of viscous stresses in FGM remains essentially empirical and the problem deserves a more specific attention. The aim of the present paper is to develop a general, rational, closure of the viscous stress tensor of the two-phase medium, and validate this closure against appropriate experimental data capable of discriminating between predictions of different models. To this end, we first analyze in Sec. \ref{sec2} the assumptions and the averaging process on which the governing equations of the one-fluid model are based. Then, we show in Sec. \ref{sec3} how the relation between the components of the stress tensor of the two-phase medium and those of the strain-rate tensor behaves in a particular two-dimensional configuration with a flat interface parallel to one of the grid axes. In Sec. \ref{sec4}, we generalize this relation to arbitrary orientations of the interface by using existing theoretical results describing the constitutive laws of anisotropic fluids. Then, in Sec. \ref{sec5}, we compare predictions obtained with the various models, including the one developed here, with original experimental data. These data are obtained by producing a buoyancy-driven exchange flow with two almost immiscible fluids having a negligible interfacial tension and a large viscosity contrast in a long vertical pipe closed at both ends (details on the experiments are provided in appendix \ref{appendex}). To compare these data with predictions of the different models, simulations making use of the VOF approach (with technical details given in appendix \ref{appendnum}) are run in a configuration replicating experimental conditions. Numerical predictions obtained on different grids (grid influence is discussed in appendix \ref{conv}) reveal that only the anisotropic model derived in Sec. \ref{sec4} is able to correctly predict the observed dynamics at a reasonable cost, and highlight the shortcomings of available \textit{ad hoc} models. Section \ref{sec6} discusses a potential numerical issue to take care of with the anisotropic model under certain severe conditions that may be encountered in counter-current flows with a large viscosity contrast. This configuration is also used to show that, apart from such `pathologic' conditions, predictions of the anisotropic model become grid-independent well before those of standard \textit{ad hoc} models, saving a large amount of computational resources. 

\section{Local vs spatially filtered one-fluid representation of two-phase flows}
\label{sec2}
In the one-fluid formulation, incompressible two-phase flows are assumed to be governed by the single set of equations 
\begin{subequations}
\label{Eq: NS}
\begin{eqnarray}
\nabla \cdot \mathbf{V} & = & 0
\label{Eq: NS1} \\
\rho \left(\partial_t \mathbf{V}+ \mathbf{V}\cdot \nabla \mathbf{V}\right)
 & = & -\nabla P+\nabla \cdot  \mathbf{T}+\mathbf{f_I}+\rho {\bf{g}}\quad   
\label{Eq: NS2}
\end{eqnarray}
\end{subequations} 
In \eqref{Eq: NS1}-\eqref{Eq: NS2}, $\mathbf{V}$ and $P$ denote the velocity and pressure fields, respectively, $\rho$ is the local density of the two-phase medium, $\mathbf{T}$ its viscous stress tensor, $\mathbf{f_I}$ is the capillary force per unit volume acting on the interface, and $\mathbf{g}$ denotes a possible external body force, such as gravity. Assuming no phase change, the shape and topology of the interface are determined by the evolution of a function $C$ governed by the transport equation
\begin{equation}
\partial_t {C}+ \mathbf{V}\cdot \nabla C = 0
 \label{Eq: C}
\end{equation} 
The exact definition of $C(\mathbf{x},t)$ depends on the specific FGM under consideration. In Front Tracking and Immersed Boundary Methods \cite*{Unverdi92, Tryggvason01, Mittal05, Griffith20, Verzicco23}, $C$ is a singular indicator function which is zero except at the interface, and is usually referred to as the interface Dirac function. In contrast, in VOF methods \cite*{Hirt81, Scardovelli99, Prosperetti07,Tryggvason11,Popinet18} $C$ 
is an indicator function characterizing the presence of one of the fluids (hereinafter chosen to be fluid 1), while in Level Set methods $C$ is a signed distance function which is zero on the interface itself \cite*{Sussman94, Sethian96, Sethian03,Gibou18}. 
The physical properties characterizing the two-phase medium at a given point $\mathbf{x}$ of the flow may be expressed with the aid of the local value of $C(\mathbf{x},t)$. In particular, the density $\rho$ is defined as
\begin{equation}
\rho=F(C)\rho_1+(1-F(C))\rho_2\,,
 \label{Eq: rho}
\end{equation}
with $\rho_1$ and $\rho_2$ being the densities of fluids 1 and 2, respectively. Similarly, if both fluids are Newtonian, the viscous stress tensor obeys the usual linear relationship $\mathbf{T}=2\mu \mathbf{S}$ everywhere, with  $\mathbf{S}=\frac{1}{2}\left(\nabla \mathbf{V}+^\text{T}\nabla \mathbf{V}\right)$ being the strain-rate tensor and 
\begin{equation}
\mu=F(C)\mu_1+(1-F(C))\mu_2\,,
 \label{Eq: mu}
\end{equation}
with $\mu_1$ and $\mu_2$ the viscosities of fluids 1 and 2, respectively. The choice of the function $F(C)$ varies from one FGM to another. In particular $F(C)=C$ in the VOF approach, while $F(C)=H(C)$ in the Level Set approach, $H$ denoting the Heaviside function. In the Front Tracking and IBM techniques, $F$ is also a Heaviside function and the norm of its gradient defines the interface Dirac function, i.e. one has $C=\left\Vert \nabla F\right\Vert$. These definitions have the attractive property that, owing to (\ref{Eq: C}), $D_t\rho=D_t\mu=0$, with $D_t=\partial_t+\mathbf{V}\cdot\nabla$ denoting the Lagrangian derivative. Hence, the density and viscosity of the medium are constant along the interface, consistent with the assumption that fluids 1 and 2 have constant properties in the absence of concentration or temperature variations.\\
\indent As long as they are considered at the level of an infinitesimal fluid parcel, (\ref{Eq: NS1})-(\ref{Eq: NS2}) are the exact Navier-Stokes equations including capillary and body forces, and (\ref{Eq: C})-(\ref{Eq: mu}) provide the exact value of the material properties of the fluid present at point $\mathbf{x}$ at time $t$. However, any numerical technique requires the equations governing the continuous problem to be discretized on a finite grid. Hence, the system of equations that is actually solved results from a spatial filtering, the size $\Delta$ of the spatial filter being typically of the same order as the grid cell size. Moreover, numerical schemes cannot deal with strict discontinuities such as those of the Heaviside and/or Dirac generalized functions involved in the above system. Therefore, it is necessary to smooth out these singular behaviors over a few layers of cells located on both sides of the interface. In most methods, the thickness of this smoothing layer (which we refer to as the transition region in what follows) ranges from $1.5\Delta$ to $2.5\Delta$ on each side of the interface \cite*{Peskin77, Unverdi92, Sussman94, Sethian96, Prosperetti07, Tryggvason11}. These filtering and smoothing processes imply that, within the transition region, the flow described by (\ref{Eq: NS})-(\ref{Eq: mu}) is actually that of a two-phase mixture the material properties of which experience steep gradients.\\
\indent Based on these remarks, and assuming a given small-but-finite $\Delta$, 
it is necessary to reconsider (\ref{Eq: NS})-(\ref{Eq: mu}) as describing a mixture, the composition of which varies continuously from that of fluid 1 to that of fluid 2 across the transition region. With this viewpoint, the above definition of each variable has to be reconsidered in the light of the filtering process. 
For instance, in the VOF approach, the spatial filter associated with the discretization is a top-hat filter of width $\Delta$. 
Therefore, the filtered value of the indicator function over the elementary control volume $\mathcal{V}_\Delta$ (i.e. the grid cell) centered at point $\mathbf{x}$ becomes the local volume fraction of fluid 1, hereinafter referred to as $\hat{C}$.
With this re-interpretation, the filtered form of (\ref{Eq: rho}) is readily shown to be the correct definition of the density of the local two-phase medium. Indeed, mass is an extensive variable, a property implying (\ref{Eq: rho}) when the density of the medium is defined as the mass enclosed in an elementary control volume $\mathcal{V}_\Delta$ with volume $V_{\Delta}$, ${M}_\Delta=[\rho_1\hat{C}{V_\Delta}+\rho_2(1-\hat{C}){V_\Delta}]$, divided by $V_{\Delta}$. In contrast, viscosity is not an extensive quantity and the foregoing argument does not apply to (\ref{Eq: mu}). Moreover the viscous stress tensor $\mathbf{T}$ involves the product of two quantities, namely the viscosity and the strain-rate tensor, both of which are discontinuous at the interface. Consequently, obtaining a correct expression for $\mathbf{T}$ requires the product of two discontinuous quantities to be correctly evaluated in any control volume straddling the interface.

\section{A two-dimensional reference configuration}
\label{sec3}
\begin{figure}
\vspace{5mm}
    \centerline{\includegraphics[scale=0.3]{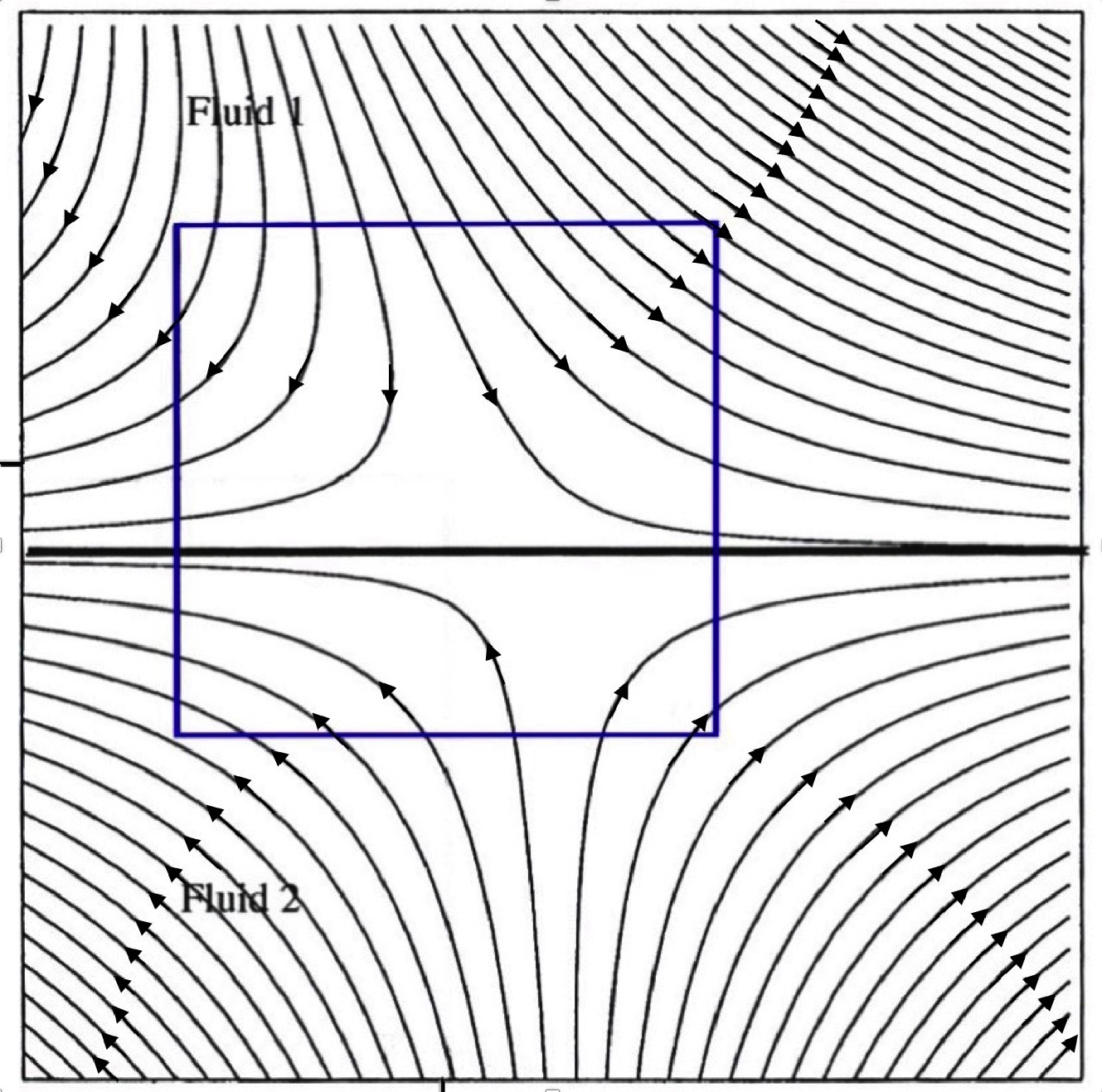}}
    \vspace{-42mm} \hspace{-81mm}
    $y=y_V$
    \vspace{2.5mm}\\
    \hspace{-83mm} $y=0$
      \vspace{0.5mm}\\
      \hspace{2mm}
    \textcolor{black}{${\mathcal{V}}_V$}
       \vspace{28mm}\\
        \hspace{-12mm}
    $x=x_V$
    \vspace{2mm}
  \caption{Sketch of the two-dimensional flow considered in Sec. \ref{sec3}. The thick black line indicates the interface while the blue square represents the elementary control volume $\mathcal{V}_V$. The thin lines depict streamlines of the flow field \eqref{Eq: vfield}.}
    \label{fig:1}
 \end{figure}

Let us now consider the behavior of the stress tensor within such an `interfacial' control volume. For the sake of simplicity, we shall make use of the definition of $\hat{C}$ and $F(\hat{C})$ suitable in the context of the VOF method, but the results to be established below remain valid with any other definition of these quantities. We consider a two-dimensional flow field with \textcolor{black}{a stationary} interface between fluid 1 and fluid 2 located at $y=0$. We define a control volume $\mathcal{V}_V$ with size $(\Delta_x,\Delta_y)$ centered at $\mathbf{x}_V\equiv(x_V, y_V)$, so that a fraction $\hat{C}=(y_V+\Delta_y/2)/\Delta_y$ (resp. $1-\hat{C}=(\Delta_y/2-y_V)/\Delta_y$) of $\mathcal{V}_V$ is filled with fluid 1 (resp. fluid 2) (see Fig. \ref{fig:1}). Within $\mathcal{V}_V$, the two components of the exact (unfiltered) velocity field in each fluid $i=1,2$ may be expanded linearly in the form 
\begin{equation}
U_{i}(x,y)=U_o+\alpha x+ \beta_{i} y\quad\textrm{,} \quad  V_{i}(x,y)=-\alpha y
 \label{Eq: vfield}
 \end{equation}
 
The velocity field (\ref{Eq: vfield}) results from the superposition of a simple shear of strength $\beta_i$ and a pure strain of strength $\alpha$. It is divergence-free and satisfies both the no-penetration condition, $V_i(x,y=0)=0$, and the continuity of tangential velocities, $U_1(x,y=0)=U_2(x,y=0)$, at the interface. Moreover, by selecting  $\beta_1$ and  $\beta_2$ such that $\mu_1\beta_1=\mu_2\beta_2$, it also satisfies the continuity of shear stresses. In the part of $\mathcal{V}_V$ filled with fluid 1, the exact viscous stress tensor is $\mathbf{T}_1=\left(\begin{array}{cc}2\mu_1\alpha & \mu_1\beta_1 \\\mu_1\beta_1 & -2\mu_1\alpha\end{array}\right)$, while the exact strain-rate tensor is $\mathbf{S}_1= \left(\begin{array}{cc}\alpha & \beta_1/2 \\\beta_1/2 & -\alpha\end{array}\right)$. Similarly, $\mathbf{T}_2=\left(\begin{array}{cc}2\mu_2\alpha & \mu_2\beta_2 \\\mu_2\beta_2 & -2\mu_2\alpha\end{array}\right) $ and $\mathbf{S}_2= \left(\begin{array}{cc}\alpha & \beta_2/2 \\\beta_2/2 & -\alpha\end{array}\right)$ in the part of $\mathcal{V}_V$ filled with fluid 2. Note that, in contrast to the off-diagonal (shear) stresses, the diagonal components of the viscous stress tensor are discontinuous at the interface as soon as $\mu_1\neq\mu_2$. This is due to the continuity of the tangential velocities, which forces the velocity gradient $\partial_x{U}_i$ to be continuous at the interface, implying by virtue of continuity the same property for $\partial_y{V}_{i}$, i.e. $\partial_y{V}_{1}|_{y=0}=\partial_y{V}_{2}|_{y=0}$. Hence, the viscosity jump translates directly into a jump of the diagonal viscous stresses. As the present example makes clear, the jump experienced by the normal viscous stresses has nothing to do with that introduced in the pressure field by capillary effects. However, one may expect this jump to play a significant role in the dynamics of certain classes of viscous-dominated two-phase flows. \textcolor{black}{The flow kinematics encapsulated in the velocity field \eqref{Eq: vfield} is sufficient to reveal the connection between all components of the filtered strain-rate tensor and those of the filtered viscous stress tensor in this two-dimensional configuration. In particular, there is no need to consider dynamical effects, since the constitutive law for the filtered viscous stress tensor is entirely determined by the local structure of the two-phase flow and cannot depend on inertial effects, hence on $\rho_1$ and $\rho_2$. Nevertheless, for the sake of completeness, it is worth 
evaluating advective terms induced by the velocity field \eqref{Eq: vfield}. Integrating \eqref{Eq: NS2} yields the pressure field in each fluid in the form $P_i(x,y)=-\rho_i[\alpha U_0x+\frac{\alpha^2}{2}(x^2+y^2)]+\rho_i[\int_0^xg_xdx'+\int_0^yg_ydy']+C_i$, with $C_i$ a constant. Selecting $C_i=-2\mu_i\alpha$ makes the total normal stress $-P_i+2\mu_i\partial_yV_i$ continuous across the interface whatever $g_x$ if the two fluids have the same density. Conversely, the body force must comprise a $x$-dependent streamwise component $g_x=\alpha[U_0+\alpha^2x]$ for this continuity to be satisfied when $\rho_1\neq\rho_2$ in this stationary configuration. In both cases, the continuity of the total normal stress across the interface implies a pressure jump $(P_2-P_1)|_{y=0}=2\alpha(\mu_1-\mu_2)$.} 
 \\
\indent From the above expressions of the strain-rate and viscous stress tensors in both fluids, the filtered viscous stress tensor of the two-phase medium may be obtained by performing a volume average of $\mathbf{T}_1$ and $\mathbf{T}_2$ over the entire control volume $\mathcal{V}_V$, which yields
\begin{equation}
\hat{\mathbf{T}}= \left(\begin{array}{cc}2(\hat{C}\mu_1+(1-\hat{C})\mu_2)\alpha & \hat{C}\mu_1\beta_1+(1-\hat{C})\mu_2\beta_2 \\\hat{C}\mu_1\beta_1+(1-\hat{C})\mu_2\beta_2 & -2(\hat{C}\mu_1+(1-\hat{C})\mu_2)\alpha\end{array}\right)\,,\quad
\hat{\mathbf{S}}=\left(\begin{array}{cc}\alpha & (\hat{C}\beta_1+(1-\hat{C})\beta_2)/2 \\(\hat{C}\beta_1+(1-\hat{C})\beta_2)/2 & -\alpha\end{array}\right)
 \label{Eq: strainrate}
\end{equation} 	
The velocity field $\hat{\mathbf{V}}$ resulting from the filtering of the exact velocity field \eqref{Eq: vfield} may also be determined at any geometrical position $\mathbf{x}\equiv(x,y)$ by averaging (\ref{Eq: vfield})  over a control volume $\mathcal{V}$ centered at $\mathbf{x}$ and having the same size $(\Delta_x,\Delta_y)$ as $\mathcal{V}_V$. If this control volume straddles the interface, i.e. if $y$ is such that $|y|\leq\Delta_y/2$, this yields
\begin{equation}
\label{filterdv}
 \hat{U}(x,y, \Delta_x,\Delta_y)=U_o+\alpha x+(2\Delta_y)^{-1}[\beta_1(y+\Delta_y/2)^2-\beta_2(y-\Delta_y/2)^2]\,,\quad \hat{V}(x,y, \Delta_x,\Delta_y)=-\alpha y\,.
 \end{equation}
 Noting that $\hat{C}(x,y, \Delta_x,\Delta_y)=\Delta_y^{-1}(y+\Delta_y/2)$ in such `mixed' control volumes, it appears that $\partial_x\hat{U}=\alpha$, $\partial_y\hat{U}=\hat{C}\beta_1+(1-\hat{C})\beta_2$ and $\partial_y\hat{V}=-\alpha$. Hence, at the specific position $y=y_V$, for which $\mathcal{V}\equiv\mathcal{V}_V$, the strain-rate tensor $\bm{\mathcal{S}}(\hat{\mathbf{V}})=\frac{1}{2}\left(\nabla \hat{\mathbf{V}}+^\text{T}\nabla \hat{\mathbf{ V}}\right)$ turns out to be identical to the filtered strain-rate tensor $\hat{\mathbf{S}}$ obtained in \eqref{Eq: strainrate}. Since $\hat{\mathbf{S}}$ was obtained by first differentiating the exact velocity field (\ref{Eq: vfield}) and then averaging over $\mathcal{V}_V$, the equality between  $\hat{\mathbf {S}}$ and $\bm{\mathcal{S}}$ illustrates the fact that the spatial derivative and filtering operators commute, even when applied to an unfiltered velocity field with discontinuous derivatives. 
The result $\bm{\mathcal{S}}\equiv\mathbf{\hat{S}}$ is important because the filtered velocity field $\hat{\mathbf{V}}$ is the central unknown that is determined in practice by solving the filtered counterpart of equations (\ref{Eq: NS1})-(\ref{Eq: NS2}). Hence, a closed momentum equation may only be obtained if the viscous stress tensor $\hat{\mathbf{T}}$ is expressible through a constitutive law involving solely $\hat{\mathbf{V}}$ and its gradients.
The above expression for $\hat{\mathbf{T}}$ may be transformed by making use of the continuity of shear stresses at the interface. The equality $\mu_1\beta_1=\mu_2\beta_2$ expressing this condition may be written in the form 
\begin{eqnarray}
\frac{\beta_1}{1/\mu_1}=\frac{\hat{C}\beta_1}{\hat{C}/\mu_1}=\frac{\beta_2}{1/\mu_2}=\frac{(1-\hat{C})\beta_2}{(1-\hat{C})/\mu_2}
=\frac{\mu_1\mu_2}{(1-\hat{C})\mu_1+\hat{C}\mu_2}\left(\hat{C}\beta_1+(1-\hat{C})\beta_2\right)\,,
 \label{Eq: stress}
\end{eqnarray} 	
 so that $\hat{\mathbf{T}}$ may be recast in the form 
\begin{equation}
\hat{\mathbf{T}}=2\left(\begin{array}{cc}\lambda\alpha & \kappa(\hat{C}\beta_1+(1-\hat{C})\beta_2)/2 \\\kappa(\hat{C}\beta_1+(1-\hat{C})\beta_2)/2 & -\lambda\alpha\end{array}\right),
 \label{Eq: stresstensor}
\end{equation} 	
with the viscosities $\lambda$ and $\kappa$ defined as
\begin{subequations}
\label{Eq: visco}
\begin{eqnarray}
\lambda(\hat{C})&=&\hat{C}\mu_1+(1-\hat{C})\mu_2\,,
\label{Eq: visclin} \\
\kappa(\hat{C})&=&\frac{\mu_1\mu_2}{(1-\hat{C})\mu_1+\hat{C}\mu_2}\,.
\label{Eq: viscquad}
\end{eqnarray}
\end{subequations} 
From (\ref{Eq: strainrate}) and (\ref{Eq: stresstensor}), it can be concluded that the components of the spatially-filtered viscous stress tensor $\hat{\mathbf{T}}$ characterizing locally the two-phase medium are linearly related to those of the strain-rate tensor $\bm{\mathcal{S}}=\frac{1}{2}(\nabla \hat{\mathbf{V}}+^T\nabla \hat{\mathbf{V}})$ resulting from the filtered velocity field. However, in contrast to the usual case of Newtonian fluids for which the relation $\mathbf{T}=\mathcal{F}(\mathbf{S})$ involves only one viscosity (in the incompressible limit), the above analysis shows that two different viscosities, $\lambda$ and $\kappa$, are required to characterize correctly the transition region in the two-phase `filtered' medium. This result is consistent with the heuristic conclusions  of \cite{Coward97}. This increased complexity results from the fact that this filtered medium is neither homogeneous (because its properties depend on $\hat{C}$ which itself depends on $\mathbf{x}$), nor isotropic in this region. Specifically, the two-phase medium is only isotropic in directions locally parallel to the interface. The above example establishes the expression of the viscosities $\lambda$ and $\kappa$ in a particular case where the interface is parallel to one of the grid coordinates. Equations (\ref{Eq: stresstensor}) and (\ref{Eq: visco}) show that in this configuration the diagonal or normal viscous stresses are proportional to a viscosity $\lambda$ varying linearly between $\mu_1$ and $\mu_2$ with the local volume fraction of each fluid. It may be noticed that (\ref{Eq: visclin}) is identical to the widely used definition (\ref{Eq: mu}) of $\mu$ in the VOF approach. In contrast, shear stresses are proportional to a viscosity $\kappa$ resulting from a harmonic average of the viscosities of fluids 1 and 2. This second result is reminiscent of the simpler case of diffusive heat transfer across the interface separating two media with different conductivities. In that case, it is well known that ensuring the continuity of heat fluxes across the interface requires the introduction of a conductivity defined as the harmonic average of the conductivities of the two media \cite*{Ferziger03, Patankar80}. The above expression for $\kappa$ could be anticipated, as the role of the shear stress in momentum transfer is similar to that of the heat flux in energy transfer.

\section{General form of the constitutive law for the filtered stress tensor}
\label{sec4}
The above results were derived in a specific configuration with a particular orientation of the interface. In order to establish the general tensorial form of the relation $\hat{\mathbf{T}}=\mathcal{F}(\bm{\mathcal{S}})$ valid whatever this orientation with respect to the grid, it is necessary to come back to the foundations of constitutive laws relating the stress tensor of a fluid (a solid) to the strain-rate tensor (the deformation tensor). When this relation is assumed to be linear, it is known to involve a fourth-order viscosity (elasticity) tensor $\mathbf{\Lambda}$ with $81$ components. In most cases, this number may be drastically reduced by invoking several arguments. Among them, some are very general as they result from invariance considerations or from the existence of a quadratic dissipation function, whereas others (especially the symmetry of the stress tensor) are related to the properties of the microstructure of the medium (see e.g. \cite{Eringen75} for a detailed discussion). Finally, the possible symmetries of the medium have to be examined in order to further reduce the number of components of $\mathbf{\Lambda}$. Employing the tools provided by the theory of representations, Ericksen \cite{Ericksen60} carried out the analysis summarized above for the fluids of interest here. In particular, he considered the case of a nonpolar, anisotropic fluid with a single preferential direction undergoing an incompressible motion. Using preliminary results for transverse anisotropic tensors obtained by Smith and Rivlin \cite{Smith57}, he showed that for such a fluid, the components of the complete stress tensor $\mathbf{\Sigma}$ obey the constitutive law
\begin{eqnarray}
\Sigma_{ij}=(a_0+a_1S_{km}n_kn_m)I_{ij}+(a_2+a_3S_{km}n_kn_m)n_in_j 
+a_4S_{ij}+a_5(S_{ik}n_kn_j+S_{jk}n_kn_i)\,,
\label{Eq: stresstensorij}
\end{eqnarray} 	
where $\mathbf{I}$ is the unit tensor, $\mathbf{n}$ is the unit normal along the preferential direction, and the six coefficients $a_0-a_5$ characterizing the fluid depend in general on time and space. Equation (\ref{Eq: stresstensorij}) reduces to the usual Newtonian constitutive law when no preferential direction exists. We may first note that the various terms in (\ref{Eq: stresstensorij}) behave differently under time reversal: since $\mathbf{V}$ transforms into $-\mathbf{V}$ in such a reversal, terms proportional to $a_0$ and $a_2$ are left unchanged while all other terms are replaced by their opposite. Hence, the former two terms account for inviscid effects while all others represent viscous contributions. In particular, $S_{km}n_kn_m=\mathbf{n}.\mathbf{S}\cdot\mathbf{n}$ is the compression/dilatation rate of the fluid in the preferential direction with unit normal $\mathbf{n}$, i.e. the one-dimensional counterpart of $\nabla \cdot \mathbf{V}$. Defining the two-dimensional divergence $\nabla_S\cdot {\mathbf{V}}=\nabla \cdot \mathbf{V}-\mathbf{n}\cdot\mathbf{S}\cdot\mathbf{n}$ and invoking the solenoidality constraint, this term may also be re-interpreted as the opposite of the surface divergence of $\mathbf{V}$. Making use of this definition and introducing the surface unit tensor $\mathbf{I}_S=\mathbf{I}-\mathbf{n}\mathbf{n}$, (\ref{Eq: stresstensorij}) may be rearranged in the form
\begin{eqnarray}
\Sigma_{ij}=[a_0+a_2-(a_1+a_3+2a_5)\nabla_S\cdot{\mathbf{V}}]I_{ij}
+[(a_3+2a_5)\nabla_S\cdot {\mathbf{V}}-a_2]I_{Sij}+\mathcal{T}_{ij}\,,
 \label{Eq: stresstensorij2}
\end{eqnarray} 	
where the traceless tensor $\bm{\mathcal{T}}$ is defined as
\begin{eqnarray}
\mathcal{T}_{ij}=a_4S_{ij}
+a_5[S_{ik}n_kn_j+S_{jk}n_kn_i+2(\nabla_S\cdot {\mathbf{V}})(I_{ij}-I_{Sij})]\,,
 \label{Eq: Tij2}
\end{eqnarray}
or equivalently
\begin{eqnarray}
\mathcal{T}_{ij}=a_4S_{ij}
+a_5[S_{ik}n_kn_j+S_{jk}n_kn_i-2(n_kS_{kl}n_l)n_in_j]\,.
 \label{Eq: Tij3}
\end{eqnarray}
The above definition of $\bm{\mathcal{T}}$ ensures that the term within square brackets does not contribute to the normal component of the viscous traction $\bm{\mathcal{T}}\cdot{\bf{n}}$ at the interface, so that ${\bf{n}}\cdot\bm{\mathcal{T}}\cdot{\bf{n}}=-a_4\nabla_S\cdot{\bf{V}}$. \\
\indent We can now specialize the above form of $\mathbf{\Sigma}$  to the one-fluid formulation of the two-phase flow equations.  First, in the context of the fluids under consideration, no three-dimensional isotropic effect due to the compression/dilatation of the interface is to be expected. Hence, no term of the form $(\nabla_S\cdot{\mathbf{V}})I_{ij}$ can exist in \eqref{Eq: stresstensorij2}, implying $a_3+2a_5=-a_1$, which in turn allows the contribution proportional to $I_{Sij}$ to be recast in the form $-[a_1(\nabla_S\cdot{\mathbf{V}})+a_2]I_{Sij}$. The first term in this contribution captures possible dilatation/compression interfacial effects preserving a two-dimensional isotropy in directions parallel to the interface. In particular, $-a_1$ may be identified as the two-dimensional counterpart of the so-called second (or volume) viscosity, and is frequently referred to as the surface dilatational viscosity \cite{Scriven60, Aris62}. This interfacial property is important in applications such as emulsification or foam stability and may be determined through specific experiments \cite{Edwards91, Kao92}. Nevertheless, interfacial rheology is irrelevant in the simpler context considered here. Therefore, we simply set $a_1=0$ in what follows. This assumption may be thought of as the two-dimensional  counterpart of the well-known Stokes hypothesis $3\lambda_N+2\mu_N=0$ in Newtonian fluids, with $\lambda_N$ and $\mu_N$ denoting the volume and shear viscosities, respectively. 
With the above two conditions, (\ref{Eq: stresstensorij2}) simplifies into
\begin{equation}
\Sigma_{ij}=a_0^*I_{ij}-a_2I_{Sij}+\mathcal{T}_{ij}\,,
 \label{Eq: stresstensorij3}
\end{equation} 	
with $a_0^*=a_0+a_2$. The form (\ref{Eq: stresstensorij3}) makes it clear that $a_0^*$ has to be identified with the opposite of the pressure $P$, while the term proportional to $a_2$ is the two-dimensional counterpart of the pressure and is thus directly related to surface tension. \\
\indent In the context of the one-fluid formulation (\ref{Eq: NS}), the capillary pressure may be expressed as $\gamma \delta_S\mathbf{I}_S$, with $\gamma$ the surface tension and $\delta_S$ the interface Dirac function \cite{Drew83}. In the framework of the VOF approach, it has been established \cite{Brackbill92} that the proper filtered counterpart of this term is obtained by defining the filtered interface Dirac function, $\delta_{S\Delta}$, as 
\begin{equation}
\delta_{S\Delta}=\left\Vert \nabla \hat{C}\right\Vert\,,
 \label{Eq: deltaS}
\end{equation}	
while the filtered unit normal $\mathbf{n}_{\Delta}$ to the interface (involved in the filtered counterpart $\mathbf{I}_{S\Delta}$ of $\mathbf{I}_S$) must be defined as 
\begin{equation}
\mathbf{n}_{\Delta}=\nabla \hat{C}/\left\Vert \nabla \hat{C}\right\Vert\,.
 \label{Eq: normal}
\end{equation}
 Expressions (\ref{Eq: deltaS})-(\ref{Eq: normal}) have the decisive computational advantage that $\delta_{S\Delta}$ and $\mathbf{n}_{\Delta}$ are smooth continuous quantities defined everywhere in the flow domain rather than singular quantities only defined at the interface. Comparing the filtered expression of the capillary pressure, $\gamma\delta_{S\Delta}\mathbf{I}_{S\Delta}$, with (\ref{Eq: stresstensorij3}) yields $a_2=-\gamma\delta_{S\Delta}$. Similarly, comparing the filtered pressure, $\hat{P}$, with the 3D-isotropic contribution in \eqref{Eq: stresstensorij3} implies $a_0^*=-\hat{P}$.\\
\indent Finally, the filtered version of the traceless tensor $\bm{\mathcal{T}}$ must be identified with the viscous stress tensor $\hat{\mathbf{T}}$ involved in the filtered momentum equation. For this purpose, we first identify $\mathbf{S}$ in (\ref{Eq: Tij2}) with the filtered strain-rate tensor $\bm{\mathcal{S}}$ defined in Sec. \ref{sec3}. We note that $\hat{\mathbf{T}}$ is traceless, as shown in a particular case by (\ref{Eq: stresstensor}). This property is general for the two-phase media considered here because $\hat{\mathbf{T}}$ results from the spatial filtering of $\mathbf{T}_1$ and $\mathbf{T}_2$ which are both traceless, since the flow of both fluids is assumed to be incompressible. Hence it is consistent to identify $\hat{\mathbf{T}}$ with the filtered version of $\bm{\mathcal{T}}$.\\
\indent As found in the particular configuration considered in Sec. \ref{sec3}, the expression for $\hat{\mathbf{T}}$ involves two viscosity coefficients instead of one in the usual single-phase case. These coefficients may be readily related to the viscosities $\lambda$ and $\kappa$ defined in (\ref{Eq: visco}) by particularizing (\ref{Eq: Tij2}) to the case of a two-dimensional flow with $\mathbf{n}_{\Delta}$ parallel to the $y$ axis as in the above example. In this configuration, the term within square brackets in (\ref{Eq: Tij2}) reduces to $S_{12}$ when $i=1$ and $j =2$ or $i=2$ and $j=1$ and is zero otherwise. Comparing this result with expressions (\ref{Eq: strainrate}) and (\ref{Eq: stresstensor}) obtained for $\hat{\mathbf{S}}\equiv\bm{\mathcal{S}}$ and $\hat{\mathbf{T}}$ in the same configuration yields 
\begin{eqnarray}
a_4=2\lambda\quad,\quad a_5=2(\kappa-\lambda)\,.
 \label{Eq: alpha}
\end{eqnarray}
	Equation (\ref{Eq: alpha}) closes the determination of the stress tensor of the two-phase medium. In summary, combining matching conditions at the interface for the different components of the filtered stress tensor in a specific configuration with the general form of the stress tensor in a fluid with a preferential direction provides the simplest consistent constitutive law for the filtered stress tensor $\hat{\mathbf{\Sigma}}$ involved in the one-fluid formulation of viscous two-phase incompressible flows. 
In intrinsic form, the result reads
\begin{eqnarray}
\hat{\boldsymbol\Sigma}=-\hat{P} \mathbf{I}+\gamma \delta_{S\Delta}\mathbf{I}_{S\Delta}+2\lambda \bm{\mathcal{S}}
+2(\kappa-\lambda)[(\bm{\mathcal{S}}\cdot\mathbf{n}_{\Delta})\mathbf{n}_{\Delta}+\mathbf{n}_{\Delta}(\bm{\mathcal{S}}\cdot\mathbf{n}_{\Delta})-2(\mathbf{n}_{\Delta}\cdot\bm{\mathcal{S}}\cdot\mathbf{n}_{\Delta})\mathbf{n}_{\Delta}\mathbf{n}_{\Delta}]\,.
 \label{Eq: Sigma_i}
\end{eqnarray}	
All terms in Eq. (\ref{Eq: Sigma_i}), including the capillary contribution, were derived through the same rational approach. Therefore, as soon as the transition region has been given a finite thickness, as in any numerical FGM, the contribution proportional to $2(\kappa - \lambda)$ in (\ref{Eq: Sigma_i}) emerges as an intrinsic part of the filtered Navier-Stokes equation. As we already mentioned, the definitions of $\delta_{S\Delta}$, $\mathbf{n}_{\Delta}$, $\lambda$ and $\kappa$ involved in (\ref{Eq: Sigma_i}) vary according to the particular FGM under consideration. In the context of VOF methods, $\delta_{S\Delta}$ and $\mathbf{n}_{\Delta}$ are defined by (\ref{Eq: deltaS}) and (\ref{Eq: normal}), respectively, while the viscosities $\lambda$ and $\kappa$ are related to $\hat{C}$ and to the viscosities $\mu_1$ and $\mu_2$ of fluids 1 and 2 by (\ref{Eq: visco}). The form (\ref{Eq: Sigma_i}) of the stress tensor associated with the definitions (\ref{Eq: visco}) of $\lambda$ and $\kappa$ ensures that the momentum balance is properly satisfied at the interface, in both the normal and tangential directions. Equations (\ref{Eq: visco}) indicate that $\lambda=\kappa=\mu_1$ when $\hat{C}=1$ ($\lambda=\kappa=\mu_2$ when $\hat{C}=0$), so that the viscous stress tensor reduces to its usual single-phase form, making (\ref{Eq: Sigma_i}) valid everywhere in the flow. In the particular case of a two-dimensional flow with an interface parallel to the $i=1$ or $i=2$ direction, the viscous part of the filtered stress tensor, say $\hat{\bm{\mathcal{T}}}$, reduces to
\begin{equation}
\hat{\bm{\mathcal{T}}}(\mathcal{S})=2\left(\begin{array}{cc}\lambda \mathcal{S}_{11}& \kappa \mathcal{S}_{12} \\\kappa \mathcal{S}_{21} & \lambda \mathcal{S}_{22}\end{array}\right)\,,
 \label{Eq: stressflat}
\end{equation} 	
which is the form that was heuristically used in \cite{Coward97}. It is also worth noting that the dissipation rate per unit volume associated with the viscous part of (\ref{Eq: Sigma_i}) is $2\lambda \mathcal{S}_{ij}\mathcal{S}_{ij}+4(\kappa-\lambda)[\mathcal{S}_{ik}\mathcal{S}_{ij}n_{\Delta k}n_{\Delta j}-(\nabla_S\cdot\hat{\mathbf{V}})^2]$. This expression may be recast in the form $2\lambda [\mathcal{S}_{ij}\mathcal{S}_{ij}-2\mathcal{S}_{ik}\mathcal{S}_{ij}n_{\Delta k}n_{\Delta j}+2(\nabla_S\cdot\hat{\mathbf{V}})^2]+4\kappa[\mathcal{S}_{ik}\mathcal{S}_{ij}n_{\Delta k}n_{\Delta j}-(\nabla_S\cdot\hat{\mathbf{V}})^2]$. Expanding the strain-rate tensor $\bm{\mathcal{S}}$ into normal and tangential components with respect to the interface, it may be proved that the two groups of terms within square brackets are positive. Therefore, the second law of thermodynamics simply requires $\lambda \geq 0$ and $\kappa \geq 0$, which is obviously satisfied by (\ref{Eq: visco}).

\section{Critical assessment of the models for $\hat{\bm{\mathcal{T}}}$: exchange flow in a vertical pipe}
\label{sec5}
To assess the performance of a specific model of the viscous stress tensor and compare predictions of the various models available, it is appropriate to select a flow configuration in which the shape, and possibly the topology, of interfaces is directly influenced by viscous stresses, with capillary effects playing a negligible role. It is also desirable that some parts of the interface are essentially submitted to a pure shearing motion while others experience an extensional flow and are therefore controlled by normal stresses. 
The buoyancy-driven exchange flow of two liquids with close densities but a large viscosity contrast in a closed vertical pipe appears to be an excellent candidate for this purpose. This configuration has aroused a sustained interest over the last three decades, especially because it is regarded as an elementary albeit relevant model of the bi-directional flow suspected to take place in the magmatic chimney of persistently degassing volcanoes. Specifically, this bi-directional flow configuration might explain the major differences (frequently by orders of magnitude) noticed between the large carbon dioxide and sulphur dioxide fluxes measured in the crater of several types of volcanoes and the modest magma flux that is actually erupted \cite{Stevenson98}. Since then, this configuration has been explored theoretically \cite{Huppert07, Kerswell11, Sweeney13}, experimentally \cite{Beckett11} and numerically \cite{Suckale18}, especially with the aim of understanding how the viscosity contrast and the initial conditions determine the flow configuration that develops once the fluids are set in motion, and which constraint sets the flow rate of each of them. Indeed, fluids may arrange in an axisymmetric or weakly eccentric core-annular configuration, with only one of them in contact with the wall, or in a side-by-side configuration, with both of them in contact with the wall, at least over a certain range of vertical positions. The latter configuration is of course more challenging, owing to its intrinsically three-dimensional nature. A simple way to produce it is to first fill a pipe with a heavy viscous liquid halfway up, and then fill the upper half with a second lighter and less viscous liquid. Provided interfacial tension is weak enough, flipping the pipe upside down \textcolor{black}{makes the fluid arrangement unstable with respect to the Rayleigh-Taylor mechanism and} initiates a side-by-side flow thanks to the asymmetric disturbance resulting from the overturn. We designed such an experiment, the characteristics of which are summarized in appendix \ref{appendex}, together with the associated protocol. Each fluid is made of a water-oil mixture, the proportions of which are adjusted to obtain the desired viscosity and density contrasts. Details regarding fluid properties and their determination are also provided in appendix \ref{appendex}. In particular, it is shown that molecular diffusion and interfacial tension are both unable to produce any discernible effect, making (\ref{Eq: NS})-(\ref{Eq: C}) with $\mathbf{f_I}=\mathbf{0}$ appropriate to simulate the flow evolution. The numerical approach selected to achieve these simulations, based on the VOF method, is summarized in appendix \ref{appendnum}. Some details about the IBM technique used to enforce the no-slip condition at the circular wall are also provided there. Simulations carried out on several grids with the parameter set on which the comparison discussed below focuses are detailed in appendix \ref{conv}. \\
  \begin{figure}
  \includegraphics[height=8.1cm]{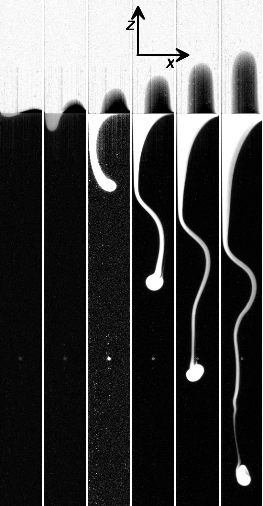}\hspace{1mm} \includegraphics[height=8.1cm]{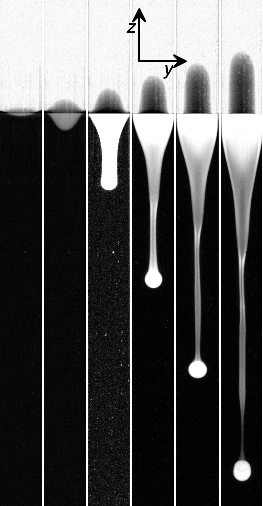}\hspace{1mm} 
  \includegraphics[height=8.1cm]{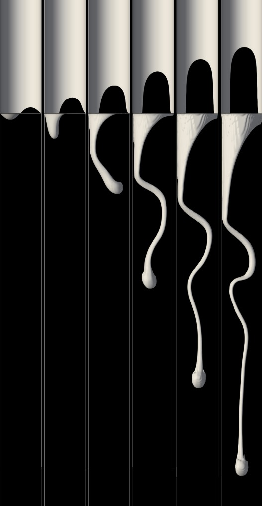}\hspace{1mm} \includegraphics[height=8.1cm]{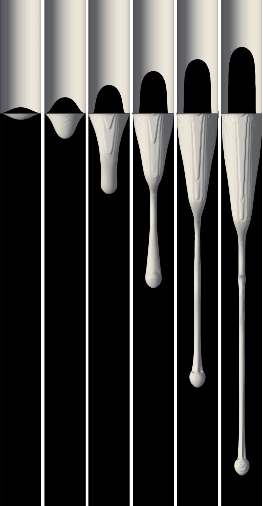}\\
        \vspace{2mm}
        \hspace{2mm}(a)\hspace{43mm}(b)\hspace{44mm}(c)\hspace{37mm}(d)
    \caption{Evolution of a buoyancy-driven exchange flow in a vertical circular pipe for a pair of fluids with $\beta\approx33, At\approx0.019$ and $Ga\approx0.39$. (a-b): front and side views of the interface in a typical experimental sequence \textcolor{black}{involving UCON oil-water mixtures with $\rho_1=1098\,$kg.m$^{-3}$, $\rho_2=1056.5\,$kg.m$^{-3}$, $\mu_1=7.0\,$Pa.s, $\mu_2=0.21\,$Pa.s}; (c-d): front and side views of  the iso-surface $\hat{C}=0.5$ in the numerical sequence based on model \eqref{Eq: Sigma_i} and a $52\times52\times
    1024$ grid (see appendix \ref{conv}). The light and heavy fluids appear in dark and light colors, respectively. Only the portion of the pipe in which both fluids are present is shown. In experiments, the flow is initiated by flipping the pipe upside down about an axis parallel to the $y$ direction and lying in the horizontal plane where the interface is located. In simulations, the iso-surface $\hat{C}=0.5$ is initially displaced vertically by imposing a disturbance vanishing at the wall and enforcing antisymmetric and symmetric $\hat{C}$-distributions in directions $x$ and $y$, respectively. Times are rescaled using the visco-gravitational time scale $\tau=\mu_1/(\rho_1-\rho_2)gD$ \textcolor{black}{and the virtual origin $t_{0d}$ defined in the text. 
    From left to right, the six snapshots in each panel correspond to rescaled times $(t-t_{0d})/\tau\approx-49, -26, -1, 30, 61$ and $91$, respectively.}}
    \label{fig:sequence}
\end{figure}
\indent To give an idea of the flow dynamics, Fig. \ref{fig:sequence} displays some snapshots showing how the interface separating the two fluids changes over time. For a given set of initial conditions, the flow evolution is entirely characterized by the viscosity ratio, $\beta\equiv\mu_1/\mu_2$, the Atwood number, $At=(\rho_1-\rho_2)/(\rho_1+\rho_2)$, and the Galilei number, $Ga\equiv[\rho_1(\rho_1-\rho_2)gD^3]^{1/2}/\mu_1$ which corresponds to the flow Reynolds number based on the gravitational velocity scale $V_g=[(1-\frac{\rho_2}{\rho_1})gD]^{1/2}$, with $D$ being the pipe diameter and $g$ denoting gravity. Throughout this section, we consider the parameter set $(\beta, At, Ga)\approx(33, 0.019, 0.39)$, for which viscous effects dominate the dynamics of the heavy fluid, and normalize time using the visco-gravitational time scale $\tau=\mu_1/(\rho_1-\rho_2)gD$. As Fig. \ref{fig:sequence} reveals, interfaces located in the upper and lower halves of the pipe follow dramatically different evolutions. In the former, the flow retains a classical core-annular structure, with a thick ascending finger of light fluid at the center, surrounded by a descending annular film of heavy fluid. Conversely, in the lower part, the heavy fluid first flows along a sector of the wall, forming a tongue that elongates over time, before bending and curving towards the pipe axis. From then on, the lower part of the descending finger gradually thins, giving rise to a long filament that continues to undulate and at the tip of which some fluid accumulates in an approximately spherical blob. In the late stages (not shown), break-up takes place and this blob turns into a separate droplet, with a new blob forming at the newborn tip. The process self-repeats until the tip gets close to the bottom of the pipe.\\
\indent Experimental results such as those of Figs. \ref{fig:sequence}(a-b) are subject to several sources of uncertainty. First, fluid viscosities are sensitive to temperature, with a typical $6\%$ change per $^\circ\text{C}$ in the case of the more viscous fluid. Therefore, even minor variations in the room temperature translate into sizable changes in the speed at which the two fluids move. Hygrometry is also important, since water evaporation modifies both the viscosity of the mixtures and their density difference. Last, as pointed out in appendix \ref{appendex}, viscosities are determined with a relative uncertainty ranging from $\pm2.5\%$ to $\pm8.5\%$, which translates into uncertainties ranging from $8\%$ to $15\%$ on $\beta$. For all these reasons, comparisons between experimental and numerical evolutions must be based on several distinct experimental runs performed with the supposedly same fluids, and only the envelope of the entire set of runs makes sense for such a comparison.\\
\indent A separate source of uncertainty lies in the lack of control on the exact shape and amplitude of the initial disturbance resulting from the finite-time overturn. In contrast, the numerical disturbance used to initiate the displacement of the interface is based on a product of sine and cosine functions that vanishes at the wall and enforces antisymmetric and symmetric distributions with respect to the vertical diametrical planes $(x,z)$ and $(y,z)$, respectively (see Fig. \ref{fig:sequence}). \textcolor{black}{Owing to the stringent lateral confinement imposed by the pipe wall, this disturbance with wavelength $D$ is the most unstable. Indeed, for two semi-infinite fluid layers with different densities and viscosities and no interfacial tension unstably superposed in a laterally unbounded domain, linear stability theory predicts that the most unstable wavelength is $\lambda_{max}=4\pi \left(\frac{\nu^2}{Atg}\right)^{1/3}$ with $\nu=\frac{\mu_1+\mu_2}{\rho_1+\rho_2}$ \citep{Mikaelian93}. The present material properties (see the caption of Fig. \ref{fig:sequence}) yield $\lambda_{max}\approx0.49\,$m, which is nearly $20$ times larger than the pipe diameter ($D=0.0264\,$m). For wavelengths $\lambda$ much smaller than $\lambda_{max}$, the disturbance grows at a rate $\omega_i(\lambda)\approx\frac{1}{4\pi}\frac{At\,g}{\nu}\lambda$, making the largest admissible wavelength $\lambda=D$ the most unstable in the present configuration. It may be noticed that $\omega_i(\lambda=D)\approx0.076\tau^{-1}$, which suggests that, in the linear regime, the initial disturbance requires approximately a $30\tau$ time period for its amplitude to grow by one order of magnitude.}\\
\indent The difference between the experimental and numerical initial disturbances makes a direct comparison of the two sets of observations at a given time irrelevant. Nevertheless, by varying the overturning time in experiments and the amplitude of the disturbance in simulations, we could conclude that details of the initial condition no longer influence the dynamics of the  descending finger once its tip has moved a $2D$-distance from the initial position of the interface. The corresponding vertical position, $z_{0d}=-2D$, and time elapsed since the start of the flow, $t_{0d}$ (which differs between experiments and simulations), may thus be employed to define a virtual origin beyond which experimental data and numerical predictions for the dynamics of this finger may be compared with confidence.  
Similar observations were made regarding the ascending finger. However, as Fig. \ref{fig:sequence} illustrates, this finger is much thicker than the descending one, which makes it motion less sensitive to the initial disturbance. For this reason, independence with respect to this disturbance is achieved earlier, when the ascending tip has moved only a $0.5D$-distance from the initial position of the interface, leading to the definition of a distinct virtual origin, $(z_{0a}, t_{0a})$, for the ascending finger.\\
\indent To achieve the comparison between experimental observations and numerical predictions, a meaningful metrics is the evolution of the vertical positions of the ascending and descending fronts, defined as the highest and lowest vertical positions at which an interface is detected (this definition includes the possible presence of detached droplets ahead of the tip of the descending filament). Figure \ref{fig:compa} shows how the position of the two fronts varies over time, and how numerical predictions based on the anisotropic model \eqref{Eq: Sigma_i}, the `arithmetic' model $\hat{\bm{\mathcal{T}}}=2\lambda(\hat{C})\bm{\mathcal{S}}$, and the `harmonic' model $\hat{\bm{\mathcal{T}}}=2\kappa(\hat{C})\bm{\mathcal{S}}$ behave. \textcolor{black}{As the discussion in appendix \ref{conv} establishes, strict grid convergence could only be achieved on the position of the ascending front with the arithmetic and anisotropic models, owing to computational costs. In contrast, non-negligible variations from one grid to another remain on the evolution of the descending finger, especially with the arithmetic and harmonic \textit{ad hoc} models. For these reasons, Fig. \ref{fig:compa} displays results obtained on three successive grids: a reference grid with $52$ cells in each horizontal direction and $512$ cells along the pipe axis (the simulated pipe is $15D$ long, which, according to tests reported in appendix \ref{conv}, is enough to simulate the flow up to the final time considered in Fig. \ref{fig:compa}), a $79\times79\times512$ grid in which the horizontal discretization is $50\%$ finer, and a $52\times52\times1024$ grid in which the vertical discretization is twice as fine.} For the aforementioned reasons, there is also a significant scatter in the slopes of experimental curves. Nevertheless, this scatter is small enough to unravel the behavior of the three models and draw clear conclusions about the quality of their predictions {\textcolor{black}{on a given grid}. It is clear at a glance that the anisotropic model \eqref{Eq: Sigma_i} is \textcolor{black}{the only one whose predictions on each grid fall within the envelope of experimental evolutions of both fronts. This success, and the failure of the two \textit{ad hoc} models on all considered grids, proves that, in this type of flow, where not only the velocity field but also the interface shape is `sculpted' by viscous effects, correct predictions may only be achieved at a reasonable cost by taking into account the anisotropy of the two-phase medium in the constitutive law for the viscous stress tensor. 
} \\
\indent Let us now examine in more detail the behavior of each of the two standard \textit{ad hoc} models. Figure \ref{fig:compa}(a) reveals that the harmonic  model severely over-predicts (by more than $35\%$ on the $52\times52\times512$ grid) the velocity of the ascending front. In contrast, the anisotropic and arithmetic models provide almost identical predictions that fall within the experimental envelope. These findings are not unlikely because, given the large viscosity contrast, the ascending finger behaves essentially as a Taylor gas bubble rising in a viscous liquid \cite{Dumitrescu43,Davies50,Goldsmith62,Zukoski66}. Continuity of shear stresses at a gas-liquid interface is known to imply vanishingly small interfacial shear stresses, leaving only significant normal viscous stresses, especially near the front stagnation point where these stresses are expected to be large, owing to the strong elongation rates in the vicinity of the dividing streamline. Therefore, only models satisfying the correct jump condition for the normal stresses can perform well, which disqualifies the harmonic model. \\
\begin{figure}
  \includegraphics[height=6.5cm]{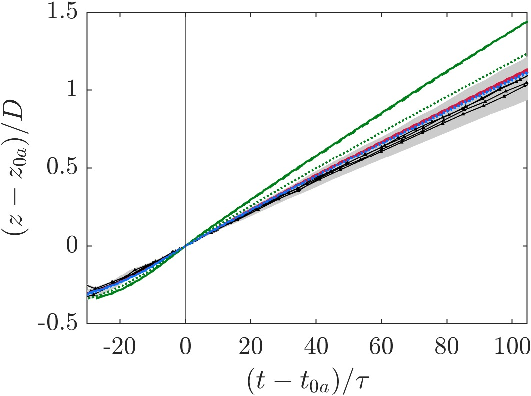}\hspace{3mm}
       \includegraphics[height=6.5cm]{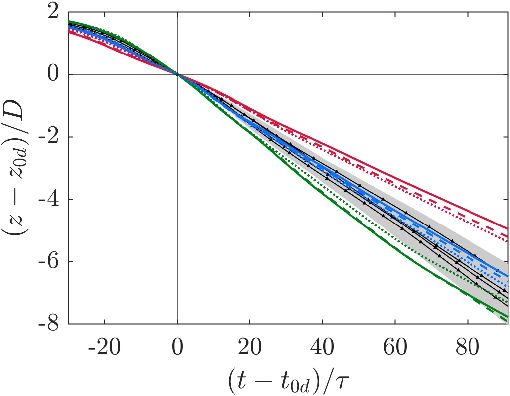}
        \vspace{2mm}
        \hspace{15mm}(a)\hspace{85mm}(b)
    \caption{Comparison of experimental evolutions of the position of the ascending and descending fronts with numerical predictions ($\beta\approx33, At\approx0.019, Ga\approx0.39$). (a): ascending front, with a virtual origin $z_{0a}=0.5D$; (b): descending front with a virtual origin $z_{0d}=2D$. Red lines: `arithmetic' model $\hat{\bm{\mathcal{T}}}=2\lambda(\hat{C})\bm{\mathcal{S}}$; green lines: `harmonic' model $\hat{\bm{\mathcal{T}}}=2\kappa(\hat{C})\bm{\mathcal{S}}$; blue lines: anisotropic model \eqref{Eq: Sigma_i}. In each series, solid, dashed and dotted lines correspond to  $52\times52\times512$, $79\times79\times512$ and $52\times52\times1024$ grids, respectively (see appendix \ref{conv} for details). \textcolor{black}{Most predictions obtained with the `arithmetic' and anisotropic models superimpose in panel (a), as do the two predictions of the `harmonic' model obtained with the lower axial resolution.} Due to the uncertainty on $\mu_1$, hence on $\tau$, the slope of all experimental series (black lines with symbols) suffers from an uncertainty ranging from $\pm2.5\%$ to $\pm8.5\%$, yielding the gray region \textcolor{black}{whose edges at a given time define the possible extreme positions of the front.}}
    \label{fig:compa}
\end{figure}
\indent Similarly, Fig. \ref{fig:compa}(b) reveals that the arithmetic model seriously underestimates (by more than $25\%$ on the same grid) the speed of the descending front. Again, this is no surprise, since the dynamics of the heavy viscous advancing tongue (or later that of the filament) are expected to be governed to a large extent by the interfacial shear stress that resists its descent. 
The arithmetic model that does not ensure the continuity of shear stresses across the interface 
cannot be expected to perform well in this situation. The figure also evidences the poor overall performance and complex behavior of the harmonic model. More specifically, this model is found to significantly overestimate (by $\approx20\%$) the speed of the descending front up to $(t-t_{0d})/\tau\approx60$, though this overestimate is less severe than in the case of the ascending front. Then, in a second stage, it predicts a gradual slowing down of the front descent, at odds with experimental curves which all exhibit an almost constant slope beyond $(t-t_{0d})/\tau\approx10$. Later, the predicted slope stabilizes at a value compatible with experimental data. These successive behaviors may be understood by first noting that, while the interface is primarily sheared by the relative motion of the heavy descending and light ascending fluids, it is also stretched, owing to the highly nonparallel nature of the  flow. This is especially true in the thick tongue that dominates the descending finger in the early stages of the flow, and still precedes the thin undulating filament in the late stages, as Figs. \ref{fig:sequence}(a-b) indicate. Since this tongue thins gradually as the distance from the initial position of the interface increases, the thinning process forces the heavy viscous fluid to accelerate spatially, imposing a significant stretching to the interface. The distribution of the stretching rate along the tongue, hence the compression rate across it, directly influences the prediction of the descent speed of the tip. However, the compression rate results in nonzero normal stresses at the interface. Since the harmonic model is unable to capture the normal stress jump across the transition region properly, it necessarily performs poorly in the first stage of the descent of the heavy fluid. In the late stage, the descent of the front is mostly controlled by the dynamics of the filament, the cross section of which reduces over time but remains fairly uniform at a given instant of time, except in the tip region where the blob forms. Due to this approximately uniform cross section, normal stresses are expected to be small over a large fraction of the interface area, making the evolution of the front essentially governed by shear stresses. This is why the harmonic model performs much better in this late stage, when it predicts a descent speed close to that found with the anisotropic model. The intermediate stage represents a transition between these two regimes, during which the predicted descent speed decreases, owing to the decreasing influence of normal stresses.\\
\indent The shortcomings of both \textit{ad hoc} models may be further specified by examining how each of them influences the viscous dissipation in the more viscous fluid. For this purpose, it must first be noticed that in the limit $\beta\gg1$, \eqref{Eq: visclin} and \eqref{Eq: viscquad} imply $\lambda(\hat{C}=\frac{1}{2})\approx\mu_1/2$ and $\kappa(\hat{C}=\frac{1}{2})\approx2\mu_2$, respectively. Hence, $\lambda/\kappa\approx\beta/4\gg1$ in the middle of the transition region. Therefore, the arithmetic model dissipates more kinetic energy in that region than the harmonic model, both through the diagonal (elongation/compression) and off-diagonal (shear) components of the strain-rate tensor. Along the ascending front, where normal stresses dominate, the harmonic model underestimates the dissipation resulting from the diagonal components of the strain-rate tensor. This makes it overestimate the descending velocity on the viscous heavy fluid in the annular film. As the overall flow rate in any horizontal cross section of the pipe is zero, this translates directly into an overestimate of the upward velocity of the light fluid, hence of the displacement of the apex of the light finger. The same reasoning applies to the descending front and explains why the harmonic model over-predicts again its speed, a consequence of the non-negligible role of normal stresses during the first half of the descent.  Also, the arithmetic model dissipates too much kinetic energy through the off-diagonal components of the strain-rate tensor, leading to an underestimate of the speed of the descending front which is, to a large extent, controlled by interfacial shear. 

\section{Limitations and strengths of the anisotropic model: the enlightening case of counter-current flows with large $\beta$}
\label{sec6}
 \begin{figure}
  \includegraphics[height=4.4cm]{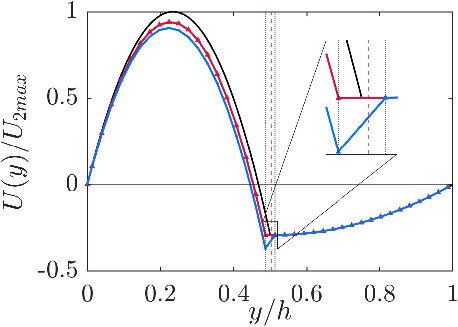}\hspace{0mm}
        \includegraphics[height=4.4cm]{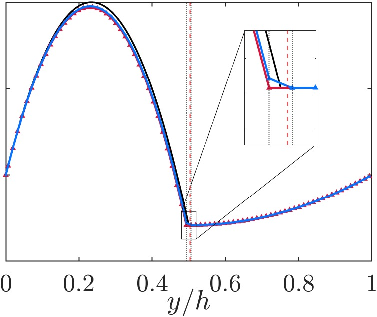}\hspace{0mm}
         \includegraphics[height=4.4cm]{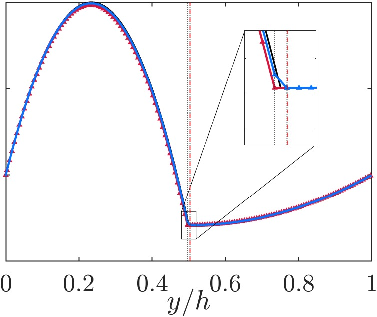}\\
        \vspace{-12.5mm}
      \hspace{11mm} Fluid 2\hspace{15mm}Fluid 1 \hspace{20mm} Fluid 2\hspace{15mm}Fluid 1 \hspace{20mm} Fluid 2\hspace{15mm}Fluid 1\\
        \vspace{10mm}
        \hspace{8mm}(a)\hspace{54mm}(b)\hspace{54mm}(c)
    \caption{A two-phase counter-current Poiseuille flow of two fluids with a large viscosity contrast ($A_2/A_1=-1/65, \beta=10^3$). (a): $\Delta/h=1/40$; (b): $\Delta/h=1/80$; (c): $\Delta/h=1/150$. The \textcolor{black}{black}, red and blue lines denote the theoretical solution (normalized by $U_{2max}$, the maximum of the velocity profile $U_2(y)$ in fluid 2)\textcolor{black}{, the prediction based on the standard model $\hat{\bm{\mathcal{T}}}=2\lambda(\hat{C})\bm{\mathcal{S}}$,} and that based on the anisotropic model \eqref{Eq: Sigma_i}, respectively. The \textcolor{black}{two} prediction\textcolor{black}{s} superimpose on the theoretical solution in the descending fluid. The vertical black dotted lines identify the boundaries of the transition region, while the red dashed line indicates the position $y=y_{m1}$ of the maximum velocity in the descending fluid.}
    \label{fig:poiseuille}
\end{figure}
The previous section established that the anisotropic model \eqref{Eq: Sigma_i} is the only one that provides correct predictions \textcolor{black}{at an affordable cost} in the flow configuration under consideration, where viscous stresses govern the interface dynamics. Nevertheless, some comments regarding a potential numerical issue that may be encountered with this model, and similarly with the \textit{ad hoc} harmonic model $\tilde{\mathbf{T}}=2\kappa(\hat{C})\mathbf{S}$, are in order. \\
\indent For the continuity of shear stresses across the interface to be satisfied once the velocity field is discretized, an obvious necessary condition is that the normal derivative of the tangential velocity keeps the same sign on both sides of the interface, i.e. the discrete value of $\partial_yU_1|_{y=y_0}\partial_yU_2|_{y=y_0}$ has to be positive if the interface is parallel to the $x$-axis such as in Fig. \ref{fig:1}. This condition is always satisfied in the case of a co-current flow, such as that encountered when the two fluids are set in motion by an external force that keeps the same sign in both of them. However, if the grid is not fine enough, it may be violated in a counter-current flow if one fluid is much more viscous than the other. To illustrate this issue, let us consider a fully-developed two-phase Poiseuille flow taking place in the gap $0\leq y\leq h$, with fluid 1 lying on top of fluid 2 and the interface being located at $y=y_0$. We assume that the flow is driven by a combination of a streamwise pressure gradient, $P_x>0$, and an external force such as gravity, with a nonzero streamwise component, $g_x>0$ (the $x$-axis is directed downwards in Fig. \ref{fig:poiseuille}). Therefore, the body force per unit volume acting on fluid 1 is $A_1=\rho_1g_x-P_x$, while that acting on fluid 2 is $A_2=\rho_2g_x-P_x$. The two velocity profiles satisfying the no-slip condition at the relevant wall are then $U_1(y)=-(A_1/2\mu_1)(h-y)^2+B_1(h-y)$ and $U_2(y)=-(A_2/2\mu_2)y^2+B_2y$. Continuity of velocities and shear stresses at the interface implies $B_1=[A_1(h-y_0)(y_0+(2\beta)^{-1}(h-y_0))+A_2\frac{y_0^2}{2}]\{\mu_1[y_0+\beta^{-1}(h-y_0)]\}^{-1}$ and $B_2=[A_1\frac{(h-y_0)^2}{2}+A_2y_0(h-y_0+\beta\frac{{y_0}}{2})]\{\mu_1[y_0+\beta^{-1}(h-y_0)]\}^{-1}$, with still $\beta=\mu_1/\mu_2$. 
The velocity profile may have extrema at positions $y_{m2}=\mu_2B_2/A_2$ and $y_{m1}=h-\mu_1B_1/A_1$ provided $0<y_{m2}<y_0$ and $y_0<y_{m1}<h$, respectively. If $\beta\gg1$, one has $y_{m2}\approx\frac{y_0}{2}$, i.e. the extremum of $U_2$ is located midway between the interface and the wall $y=0$ whatever the driving forces $A_1$ and $A_2$ \textcolor{black}{(this is no surprise since fluid 1 is almost at rest given the large viscosity contrast)}. In the same limit, the possible extremum of $U_1$ is located at a position $y_{m1}$ such that $y_{m1}-y_0 \approx -(A_2/A_1)\frac{y_0}{2}$. This extremum exists \textcolor{black}{in the gap $[y_0,h]$} only if $y_{m1}-y_0$ is positive and smaller than $h-y_0$, which happens only if $A_1$ and $A_2$ have opposite signs (a necessary condition to generate a counter-current flow) and their magnitude is such that $|A_2/A_1|<\frac{2(h-y_0)}{y_0}$. If both conditions are satisfied, the grid must capture the change of sign of $\partial_yU_1$ at $y=y_{m1}$ for $\partial_yU_1$ and $\partial_yU_2$ to have the same sign on the interface. This may become an issue if $A_2$ is small and/or $A_1$ is large, for instance if the positive density difference $\rho_1-\rho_2$ is large enough for the condition \textcolor{black}{$\rho_2g_x\lessapprox P_x\ll\rho_1g_x$} to be fulfilled. 
\textcolor{black}{In such a case, the upward motion of fluid 2 is driven by the small negative net force $A_2=\rho_2g_x-P_x$, resulting in a moderate shear rate $\partial_yU_2|_{y=y_0}$ that requests only a tiny positive shear rate $\partial_yU_1|_{y=y_0}$ in fluid 1 for the shear stress to remain continuous at the interface. This is why the velocity gradient in fluid 1 changes sign very close to the interface,} possibly making  $y_{m1}-y_0$ become of the order of the thickness of the transition region, $\Delta_\text{T}$, or even smaller. For the grid to capture the change of sign of $\partial_yU_1$ at $y=y_{m1}$, it is necessary that the extremum of $U_1(y)$ lies outside the transition region, which requires $2(y_{m1}-y_0)$ to be larger than $\Delta_\text{T}$, implying $\Delta_\text{T}/h<(-A_2/A_1)y_0/h$.\\
\indent Figure \ref{fig:poiseuille} illustrates this issue in the case of two fluids with a large viscosity contrast, $\beta=10^3$. \textcolor{black}{Due to the algorithm used to track the interface (see appendix \ref{appendnum}), the transition region reduces to a single grid cell, i.e. $\Delta_\text{T}=\Delta$.} 
With the selected driving forces, the change of sign of the velocity gradient in the more viscous fluid takes place very close to the interface since ($(y_{m1}-y_0)/h=1/130$. The coarsest grid ($\Delta/h=1/40$) is unable to capture this change of sign. Therefore the continuity of shear stresses as expressed in \eqref{Eq: stress} cannot be satisfied, although the use of the viscosity $\kappa(\hat{C})$ in the model implicitly assumes it is. This conflicting situation results in the occurrence of a `tooth' in the solution, as Fig. \ref{fig:poiseuille}(a) evidences. Because of this spurious behavior, the anisotropic model, which in the present case reduces to $\hat{\mathcal{T}}_{12}=2\kappa(\hat{C})\mathcal{S}_{12}$, performs poorly in the ascending fluid. The condition $\Delta/h<(-A_2/A_1)y_0/h$ is not yet satisfied with $\Delta/h=1/80$. However, the change of sign of $\partial_yU_1$ takes place close enough to the edge of the transition region for the discretized velocity profile to capture the associated minimum of $U_1(y)$ at the nearby location $y=y_0+\Delta/2$, removing the `tooth' from the profile predicted by \eqref{Eq: Sigma_i} (Fig. \ref{fig:poiseuille}(b)). Last, with $\Delta/h=1/150$, $\partial_yU_1$ changes sign slightly outside the transition region. Model \eqref{Eq: Sigma_i} now closely follows the theoretical solution in both fluids (Fig. \ref{fig:poiseuille}(c)). Interestingly, a behavior similar to that noticed in Fig.  \ref{fig:poiseuille}(a) was observed in \cite{Suckale18} in a counter-current flow of two miscible fluids, assuming an exponential viscosity variation in the form $\mu(\hat{C})=\mu_2\text{exp}\left[{\text{ln}\left(\frac{\mu_1}{\mu_2}\right)\hat{C}}\right]$. Authors argued that, since the combination of a linear density profile obeying \eqref{Eq: rho} and an exponential viscosity profile implies the existence of a thin layer of relatively heavy fluid with a relatively low viscosity, this layer is ``actively sinking''. However, examination of their results (figure S4 in their supplemental material, corresponding to $\beta=1.7\times10^3$) evidences that the `tooth' extends throughout the transition region $0<\hat{C}<1$ and is not limited to the sublayer $\hat{C}\lesssim1$. This indicates that a grid-related issue similar to that described above is present in their simulations. \\
 \begin{figure}
  \includegraphics[height=4.4cm]{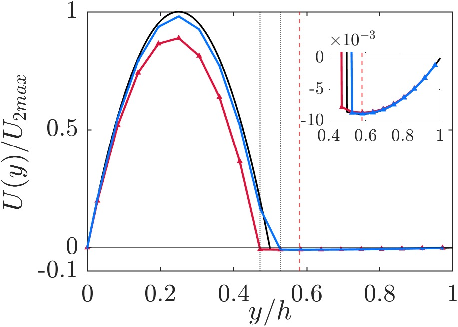}\hspace{0mm}
  \includegraphics[height=4.4cm]{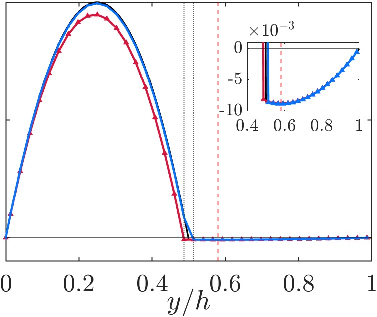}\hspace{0mm}
  \includegraphics[height=4.4cm]{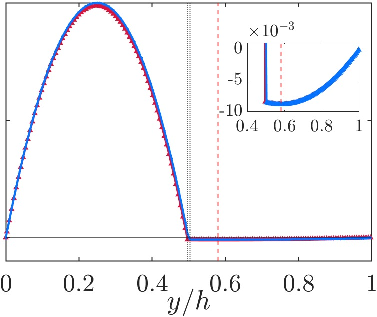}\hspace{0mm}\\
        \vspace{-15.5mm}
      \hspace{11mm} Fluid 2\hspace{15mm}Fluid 1 \hspace{20mm} Fluid 2\hspace{15mm}Fluid 1 \hspace{20mm} Fluid 2\hspace{15mm}Fluid 1\\
        \vspace{13mm}
        \hspace{8mm}(a)\hspace{54mm}(b)\hspace{54mm}(c)
    \caption{Same as Fig. \ref{fig:poiseuille} with $A_2/A_1=-0.32$. (a): $\Delta/h=1/20$; (b): $\Delta/h=1/40$; (c): $\Delta/h=1/160$. \textcolor{black}{The inset provides a detailed view of the profile of the tiny negative velocities in fluid 2. In panel (c), the numerical prediction based on the anisotropic model (blue line) superimposes on the theoretical solution.}}
    \label{fig:poiseuille2}
\end{figure}
 \begin{figure}
  \includegraphics[height=6.4cm]{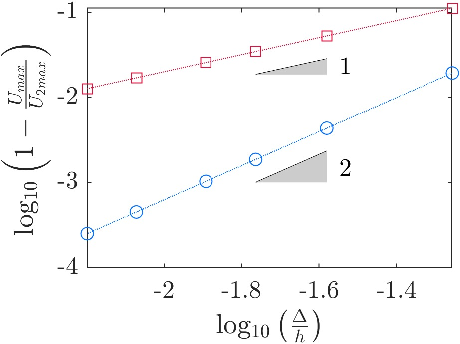}\hspace{0mm}
    \caption{\textcolor{black}{Grid convergence of the anisotropic and arithmetic models in the configuration considered in Fig. \ref{fig:poiseuille2}.}}
    \label{fig:poiseuille3}
\end{figure}
\indent Predictions of the standard arithmetic model $\hat{\mathcal{T}}_{12}=2\lambda(\hat{C})\mathcal{S}_{12}$ are also reported in Fig. \ref{fig:poiseuille}. They may give the impression that this model performs better than or at least equally well as the anisotropic model in a counter-current flow. This is actually only true when the condition $\Delta_\text{T}/h<(-A_2/A_1)y_0/h$ is violated, as in panel (a). In panels (b-c), the standard model provides accurate predictions because the grid is very fine and the transition from fluid 1 to fluid 2 is restricted to one grid cell ($\Delta_\text{T}=\Delta$). \textcolor{black}{Differences between predictions of the two models would become more pronounced if several cells were lying in the transition region. Indeed, the standard arithmetic model over-estimates the viscosity of the two-phase mixture throughout this region, therefore underestimating the velocity gradient over all cells lying therein. Consequently, the error on the velocity gradient cumulated throughout the transition region would then make the velocity in fluid 2 decrease compared to the configuration with $\Delta=\Delta_\text{T}$, resulting in a larger difference with the theoretical solution.} Differences between the two models also become much more prominent with increasing $|A_2/A_1|$, as in the example displayed in Fig. \ref{fig:poiseuille2}, where $|A_2/A_1|$ is $20$ times larger than in the previous case. In Fig. \ref{fig:poiseuille2}(a), the anisotropic model is seen to already produce accurate predictions on a coarse grid with only $20$ cells across the channel width, while the standard model under-predicts the maximum velocity $U_{max}$ and flow rate by $11\%$ and $17\%$, respectively. Predictions of the anisotropic model virtually superimpose on the theoretical solution with $40$ cells (Fig. \ref{fig:poiseuille2}(b)), while the arithmetic model still exhibits a sizable departure. \textcolor{black}{As expected, this departure decreases as the resolution increases but is still $1.3\%$ on the maximum velocity with $160$ cells (Fig. \ref{fig:poiseuille2}(c)), to be compared with the $0.4\%$ difference found with only $40$ cells in the case of the anisotropic model. 
\textcolor{black}{The rate at which predictions provided by the two models converge towards the theoretical solution is established in Fig. \ref{fig:poiseuille3}. It is seen that those based on the anisotropic model converge quadratically with the cell size, whereas those based on the arithmetic model only converge linearly. This is no surprise, as the normalized error made on the velocity gradient in the transition region is of $\mathcal{O}(\Delta_\text{})$ with the former model (since it satisfies the continuity of shear stresses at the interface at first order in $\Delta$), while it is of $\mathcal{O}(1)$ with the latter, implying errors of $\mathcal{O}(\Delta_\text{}^2)$ and $\mathcal{O}(\Delta_\text{})$ on the velocity field, respectively.
Therefore,} in agreement with the findings of the previous section, the counter-current Poiseuille flow configuration allows us to confirm that, except under `pathologic' conditions such as those of Fig. \ref{fig:poiseuille}, simulations making use of the anisotropic model (which in this specific flow configuration reduces to the standard harmonic model) converge much faster than those employing the arithmetic model, providing accurate predictions on much coarser grids, hence saving a large fraction of computational time. Note that in this example, grid convergence is easily achieved with the anisotropic model, although the viscosity ratio is very large. This contrasts with the exchange flow of Sec. \ref{sec5} where significant grid effects are still present on the finest grids considered. The reason is that, in the present configuration, viscous effects alter the velocity field but have no influence on the position an shape of the interface, whereas they control these features, especially the detailed geometry of the thin descending finger, in the exchange flow. 
}\\
\indent 

\section{Summary}
\label{sec7}
In this paper, we revisited the question of the proper modeling of viscous effects in the one-fluid formulation of incompressible two-phase flows on which all fixed-grid methods routinely used nowadays to compute complex two-phase flows are grounded. The difficulty arises because of the spatial filtering inherent to the discretization process, which, in all such methods, implies the existence of `mixed' grid cells in the transition region that connects the two `pure' fluids. A simple two-dimensional flow configuration allowed us to establish the proper relation between the filtered viscous stress tensor and the filtered strain-rate tensor in a control volume straddling the interface. This specific example makes it clear that the proper representation of the jump (continuity) of the diagonal (off-diagonal) components of the viscous stress tensor across the interface requires the introduction of two distinct viscosity coefficients instead of a single one in a usual Newtonian fluid. These two viscosities are related to those of the two pure fluids through two different averaging formulas involving the function characterizing the local presence of each fluid, i.e. the volume fraction in the case of the Volume of Fluid method. Then, available results for the stress tensor of an anisotropic fluid medium with a single preferential direction allowed us to derive the generic constitutive law for a Newtonian two-phase medium. This constitutive law, expressed in intrinsic tensorial form, is valid whatever the local orientation of the interface with respect to the grid. The unknown coefficients involved are readily determined with the help of the results obtained in the aforementioned specific configuration. \\
\indent To assess the performance of this model, together with those of available simpler \textit{ad hoc} models, we made use of a specially designed experiment in which a buoyancy-driven exchange flow is generated in a long pipe closed at both ends. By selecting almost immiscible fluids with negligible interfacial tension and a large viscosity contrast, we could obtain experimental data that, despite a significant experimental scatter, may be considered as the `justice of the peace' of all models of the viscous stress tensor. We could then compare the experimentally recorded evolution of the fronts of the ascending and descending fluids with numerical predictions \textcolor{black}{obtained on several grids with different refinements. Although no strict grid convergence was reached on the descending front,} this comparison establishes unambiguously that the anisotropic model derived here on rigorous grounds is the only one capable of reproducing experimental observations \textcolor{black}{at a reasonable cost. On the same grids,} the harmonic \textit{ad hoc} model turns out to be unable to provide realistic predictions for both the ascending and descending fronts. Its popular arithmetic counterpart also fails to predict quantitatively the dynamics of the descending front. We interpreted these shortcomings by examining the viscous dissipation each model produces in the transition region, considering separately the contribution of the diagonal and off-diagonal components of the strain-rate tensor, hence those of the normal and tangential viscous stresses, to this dissipation. Last, we examined in some detail the behavior of the anisotropic model in counter-current flow configurations with large viscosity contrasts. We showed that the numerical implementation of the model requires some caution in the selection of the minimum cell size when the body forces driving the motion of the two fluids are very unequal. We also took advantage of this configuration to \textcolor{black}{confirm} that the anisotropic model provides in almost all cases a significantly more accurate solution than the arithmetic model \textcolor{black}{on a given grid}.  Similar conclusions could certainly be reached by comparing predictions of the anisotropic and harmonic models in appropriate canonical configurations. Clearly, the reason for this success lies in the fact that the anisotropic model encapsulates the right matching/jump conditions at the interface for both tangential and normal stresses, which is not the case with any of the standard \textit{ad hoc} models. \textcolor{black}{Of course, all models are expected to converge towards the same solution as the cell size goes to zero. In that respect, adaptive mesh refinement (AMR) strategies may allow the \textit{ad hoc} models to provide fairly accurate predictions at a reasonable cost, even in complex viscosity-driven flows such as the exchange flow considered here. Nevertheless, even codes equipped with AMR would benefit from the introduction of the anisotropic model, since this model would improve their predictions at a given level of grid refinement in regions crossed by the interface.} \\
\indent Conclusions of the present work are expected to be taken into account in numerical codes aimed at simulating two-phase incompressible flows with large viscosity contrasts, as they could significantly improve the accuracy of their predictions and/or reduce the required computational resources. More specifically, only minor differences between predictions of the anisotropic model and those of the popular arithmetic model are expected in configurations where interfacial shear stresses play a secondary role, such as buoyancy-driven bubbly flows. Similarly, the \textit{ad hoc} harmonic model is presumably almost as accurate as the anisotropic model in shear-driven flows with an almost flat interface, such as in the very early stages of the development of interfacial waves. In contrast, only the anisotropic model appears capable of predicting accurately at a reasonable cost strongly non-parallel shear-driven flows with large viscosity contrasts, such as large-amplitude wind-driven capillary-gravity waves, the viscosity-driven break-up of liquid jets discharging into another liquid, not to mention breaking waves and the myriad of bubbles and droplets they produce. 

\acknowledgements
We thank G. Dhoye and J. D. Barron who designed the experimental set-up, S. Cazin for his help with the tuning of the optical system and image processing software, and P. Elyakime for his support with the JADIM code. This work was granted access to the HPC resources of the CALMIP supercomputing center under allocation 2024-[1525] and those of the CINES supercomputing center under allocation AD012A16035.
\appendix
\section{Experimental details}
\label{appendex}
\begin{figure}[h]
        \centering
        \includegraphics[height=6cm]{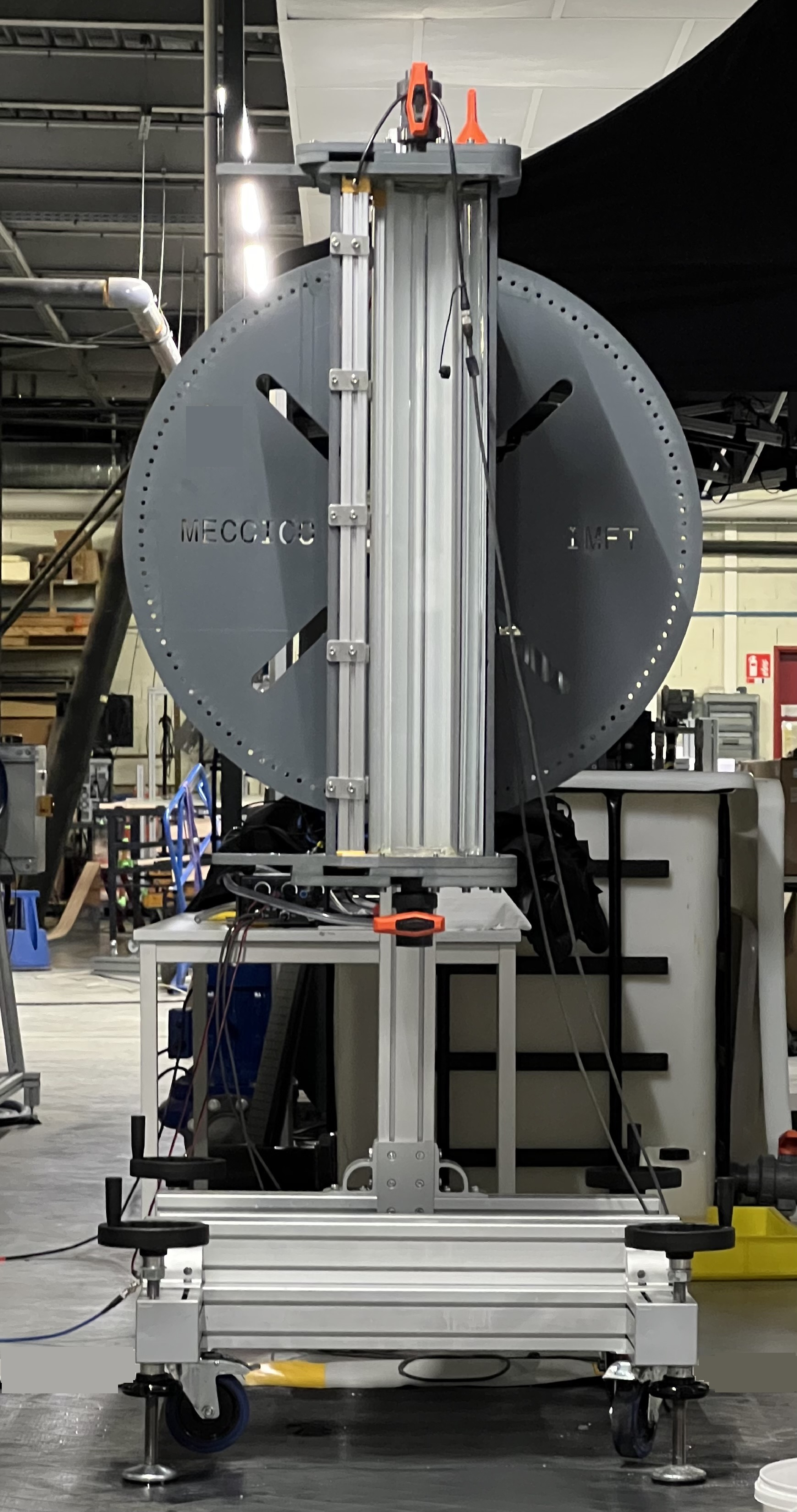}\hspace{5mm}
        \includegraphics[height=6cm]{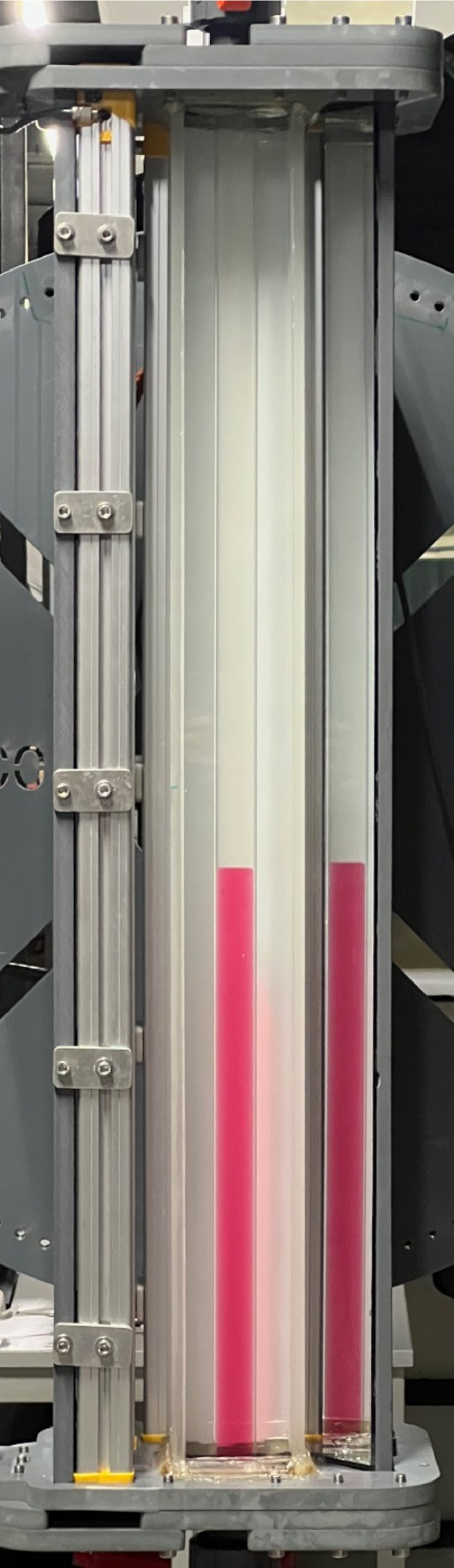}\\
        \vspace{2mm}
        \hspace{9mm}(a)\hspace{25mm}(b)
    \caption{Experimental set-up. (a): general view showing the fixed frame and the mobile assembly; the large gray disk at the back allows the pipe inclination to be varied in $3^\circ$-steps to study non-vertical configurations. (b): close-up showing a front view of the cylindrical pipe (with the heavy fluid in the bottom part colored red) surrounded by the square reservoir. On the right, the vertical mirror placed at $45^\circ$ from the reservoir provides a side view of the flow. }
    \label{fig:meccico}
\end{figure}
The key part of the device is a $1\,$m long $2.64\,$cm inner-diameter cylindrical glass pipe closed at both ends with quarter-turn taps. This pipe is surrounded by a $9.7\,$cm\,$\times\,9.7\,$cm square glass reservoir filled with water, aimed at minimizing optical distortions (see Fig. \ref{fig:meccico}(b)). The whole assembly is mounted on a frame that can be rotated manually around a horizontal shaft located midway between the pipe ends. This shaft is fixed to a vertical beam anchored in a horizontal square frame, the four feet of which may be adjusted to align precisely (with a digital inclinometer) the pipe axis with the vertical (Fig. \ref{fig:meccico}(a)). The pipe is illuminated with two $1\,$m long vertical LED backlights placed at right angle parallel to the walls of the square reservoir. Finally, a $1\,$m long vertical front-face mirror is placed at $45^\circ$ with respect to this reservoir, so that two perpendicular views of the pipe may be obtained with a single camera (Fig. \ref{fig:meccico}(b)). The flow is recorded with a PCO pco.edge 4.2 video camera with a resolution of $2048\times2048$ pixels. The camera is positioned at a distance of $10\,$m from the device to obtain the desired field of view. It is equipped with a Nikon ED - AF Micro Nikkor $200\,$mm 1:4 D lens with a specific extra-low dispersion treatment minimizing optical distortion of image edges. The acquisition rate is varied from one to three images per second, depending on the fluid viscosities. Backlights are synchronized with the camera and operate in intermittent mode, emitting $4\,$ms long flashes at the same frequency as the camera to limit fluid heating. Recorded images are post-processed with the Camware software developed for PCO cameras. A background image is first subtracted from each raw image. Once the resulting image has been binarized, a suitable filter is applied to maximize the contrast between the two fluids, allowing the interface to be eventually detected.\\
\indent Fluids used to produce the exchange flow are obtained by mixing in suitable proportions water with UCON\texttrademark\, 75-H-90,000 oil from Dow Chemical. Agitation is maintained until a homogeneous mixture is obtained. This mixture is kept at rest for one to two days, allowing bubbles trapped during the agitation process to leave through the free surface and the heat released by the oil-water chemical reaction to be evacuated. Once the two liquids are ready, the heavy one is gradually introduced at the bottom of the pipe through a flexible hose connected to a reservoir. Then, the light liquid is introduced from the top of the pipe through a second flexible hose ending just above the free surface of the heavy liquid, to minimize disturbances at the interface. The temperature of each fluid is recorded and a sample is taken to measure their respective density and viscosity. Last, the moving assembly is flipped upside down manually in typically $5\pm2$ seconds, a viscosity-dependent compromise resulting from the need to avoid any significant mixing at the interface and to limit the amplitude of the flip-induced disturbance. The pipe is eventually locked in the desired position by a stop located on the large PVC disk visible in Fig. \ref{fig:meccico}(a).\\
\indent Viscosities are determined with a Haake Mars III rheometer from Thermo Fischer Scientific, imposing the temperatures previously recorded during the filling process. It was noticed that temperature has a significant influence on the viscosity of the water-oil mixture. For instance, with a $67\%$ mass fraction of UCON oil, which corresponds to the heavy fluid used in the experiments reported in Sec. \ref{sec5}, viscosity varies by approximately $6.3\%$ per $^\circ$C in the range $[15-20]^\circ\text{C}$. The final value retained for the viscosity is based on several series of measurements performed with increasing and decreasing shear rates. An uncertainty ranging from $\pm2.5\%$ to $\pm8.5\%$ exists among the different series, resulting in a $8\%$ to $15\%$ uncertainty on the viscosity ratio $\beta$. Assumptions of immiscibility and negligible capillary effects were assessed in the following manner. After recording the growing thickness of the interfacial region in a static water-UCON oil arrangement for two days, it was concluded that the molecular diffusion coefficient is around $1\times 10^{-10}\,\text{m}^2\,\text{s}^{-1}$ in this two-phase system. As the total duration of an experimental run ranges from $5\,$min to $25\,$min, depending on the viscosity of the heavier fluid, the final penetration depth is estimated to be less than half a millimeter, i.e. less than $2\%$ of the pipe diameter. For this reason, it is appropriate to consider the two fluids as immiscible over the time range relevant for comparison with simulations. Interfacial tension was estimated using a Kr{\"u}ss FM40 tensiometer based on the pendent drop technique. For the pair of fluids with $\beta\approx33$ considered in Sec. \ref{sec5}, it was concluded that the interfacial tension does not exceed $1.5\times10^{-5}\text{Nm}^{-1}$. \textcolor{black}{Using the gravitational velocity scale $V_g=[(1-\frac{\rho_2}{\rho_1})gD]^{1/2}$ (also involved in the definition of $Ga$), the capillary number $Ca=\mu_1V_g/\gamma$ is found to be approximately $4.6\times10^4$,} confirming the absence of any significant capillary effect in this two-fluid system.
\section{Computational aspects}
\label{appendnum}
Computations reported in Sec. \ref{sec5} and appendix \ref{conv} were carried out with the in-house code JADIM. Details on the spatial discretization and time-advancement algorithm used to solve the Navier-Stokes equations \eqref{Eq: NS1}-\eqref{Eq: NS2} for $(\mathbf{\hat{V}},\hat{P})$ may be found in \cite*{Calmet97}. Briefly, the momentum equations are discretized on a staggered grid using a finite volume approach. Spatial derivatives are approached using second-order centred differences. Time-advancement is achieved through a third-order Runge-Kutta algorithm for advective and source terms and a Crank-Nicolson algorithm for `standard' viscous terms, i.e. those involved in homogeneous isotropic Newtonian fluids. Terms proportional to $2(\kappa - \lambda)$ in \eqref{Eq: Sigma_i} are treated explicitly. The transport equation \eqref{Eq: C} for the volume fraction $\hat{C}$ is split into successive one-dimensional steps along each grid direction $i$ ($i=1,3$) and solved using a Zalesak scheme \cite{Zalesak79} within each substep; see \cite{Bonometti07} for an extensive description of the transport strategy used to advance $\hat{C}$.\\
\indent
The time-advancement procedure employed to solve the coupled system \eqref{Eq: NS}-\eqref{Eq: C} starts with solving (\eqref{Eq: C}, prior to solving the momentum equation. Hence, starting from $\hat{C}^n(\mathbf{x})$ and $\mathbf{\hat{V}}^n(\mathbf{x})$ at time $n\Delta t$ (with $\Delta t$ being the time step), the solution $\hat{C}^{n+1}(\mathbf{x})$ corresponding to time $(n+1)\Delta t$ is determined first, and is employed to evaluate the density $\rho^{n+1}(\mathbf{x})=\hat{C}^{n+1}(\mathbf{x})\rho_1+(1-\hat{C}^{n+1}(\mathbf{x}))\rho_2$,
 and the viscosities $\lambda^{n+1}(\mathbf{x})$ and $\kappa^{n+1}(\mathbf{x})$ using (\ref{Eq: visclin}) and (\ref{Eq: viscquad}), respectively. Then, the second-order approximations of $\rho$, $\lambda$ and $\kappa$ at time $(n+1/2)\Delta t$ are defined as  $\rho^{n+1/2}=(\rho^n+\rho^{n+1})/2$, etc, and are used throughout the time step $[n\Delta t, (n+1)\Delta t]$ to solve (\ref{Eq: NS2}) and obtain a provisional velocity field $\mathbf{\hat{V}}^{*}$. As the volume fraction is defined at pressure nodes, linear interpolations are employed to evaluate the density and the two viscosities at velocity nodes. Finally, the velocity field is made divergence-free through a variable density projection technique. For this purpose, a pseudo-Poisson equation of the form $\nabla \cdot (\nabla\Phi^{n+1/2}/\rho^{n+1/2})=\nabla.\mathbf{\hat{V}}^{*}$ is solved to determine the pressure increment $\Phi^{n+1/2}$, from which the divergence-free velocity field is obtained as $\mathbf{\hat{V}}^{n+1}=\mathbf{\hat{V}}^{*}-(\Delta t/\rho^{n+1/2})\nabla \Phi^{n+1/2}$, whereas the pressure field is updated in the form $\hat{P}^{n+1/2}=\hat{P}^{n-1/2}+ \Phi^{n+1/2}$. 
 \\
\indent Given the pipe geometry used in experiments, a cylindrical grid is \textit{a priori} appropriate for simulating the exchange flow. Indeed, we initially ran computations on this type of grid. However, cells located close to the pipe axis have a very small circumferential length. Moreover, the circumferential velocity component takes significant values in some flow regions, especially in the descending finger. Because of these two factors, the Courant-Friedrichs-Lewy condition only allows very small time steps in this coordinate system. Given the long physical duration of experimental runs, computations must also be carried out over long periods of time. For this reason, the time step constraint imposed by cylindrical grids makes most simulations exceedingly expensive. To circumvent this difficulty, we switched to a Cartesian grid, taking the circular geometry of the wall into account with the aid of the IBM approach described in \cite{Bigot14,Pierson18}. Designed primarily for tracking rigid bodies translating and rotating in a fluid, this IBM variant is of the body-force type. Since the wall is fixed in the present case, the no-slip condition is simply enforced by adding a force density $\mathbf{f}_\text{IBM}=-\zeta\rho{\bf{\hat{V}}}/\Delta t$ to the right-hand side of \eqref{Eq: NS2}, with $\zeta({\bf{x}})$ being the local volume fraction of the solid in the computational cell at position ${\bf{x}}$. Following \cite{Pierson18}, $\zeta$ is varied from $1$ to $0$ with a sine distribution over three grid cells, with $\zeta=\frac{1}{2}$ corresponding to the desired wall position. 
Detailed tests were carried out to assess the quality of the approximation provided by this IBM approach in the near-wall region. For this purpose, two sets of computations covering two orders of magnitude of the viscosity ratio were run in the same exchange flow configuration with $At\approx0.024$ and $Ga\approx1$, one on a cylindrical grid, the other on a Cartesian grid. In the former case, the pipe cross section was discretized with $34\times32$ uniform cells in the radial and circumferential directions, respectively. The square horizontal cross section of the Cartesian grid, made of $52\times52$ uniform cells, had a side $15\%$ larger than the pipe diameter, to make sure that the grid extends well within the wall. Computations were initialized with a symmetric (bump) disturbance of the interface, leading to a core-annular configuration. Evolutions of the tip position and cross-sectional velocity profiles predicted with the two different grid systems were found to almost superimpose whatever the viscosity ratio, with, for instance, less than $1\%$ difference on the tip velocity for $\beta=100$ \citep{Mer19}. In these test cases, the Cartesian grid allowed the time step to be five times larger than with the cylindrical grid, demonstrating the interest of the Cartesian option. Based on these tests, a $15\%$ overlap between the grid and the solid region surrounding the pipe was applied in all simulations discussed in Sec. \ref{sec5} and appendix \ref{conv}.
\section{Characteristics of simulations and grid sensitivity}
\label{conv}
 \begin{figure}
  \includegraphics[height=4.4cm]{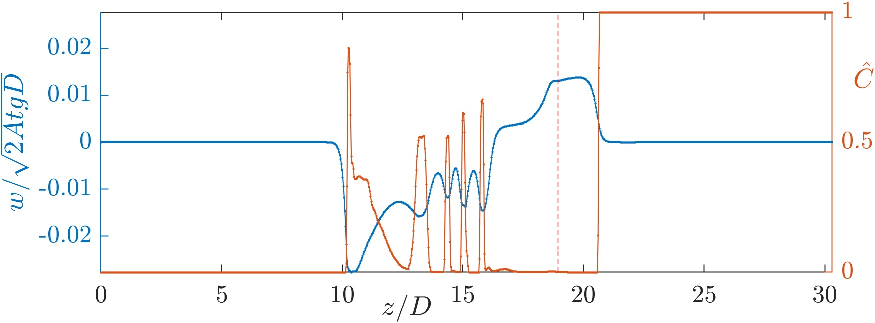}\vspace{3mm}\\
        \includegraphics[height=4.4cm]{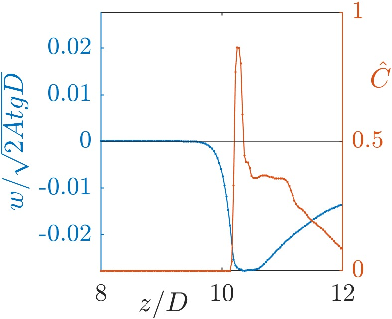}\hspace{5mm}
         \includegraphics[height=4.4cm]{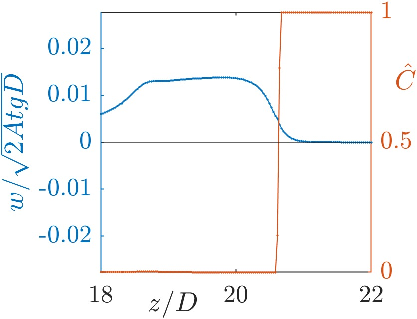}\\
        \vspace{-50mm}
        \hspace{8mm}(a)\\
         \vspace{46mm}
        \hspace{4mm}(b)\hspace{55mm}(c)
    \caption{Longitudinal variations of the axial velocity along the pipe axis at $t/\tau=150$ predicted by model \eqref{Eq: Sigma_i} on a $52\times52\times1024$ grid in a pipe with $L/D=30.3$. (a): complete distribution, from the bottom end, $z/D=0$, to the top end, $z/D=30.3$; (b) and (c): zoom around the position of the descending and ascending fronts, respectively. Blue line and left axis: axial velocity; orange line and right axis: volume fraction of heavy fluid. The vertical dashed line in (a) corresponds to the initial position of the interface.}. 
    \label{fig:axial}
\end{figure}
To assess the respective performance of the viscous stress models against experimental results, we ran simulations with the two familiar \textit{ad hoc} models and the anisotropic model \eqref{Eq: Sigma_i} in the buoyancy-driven exchange flow described in Sec. \ref{sec5} ($\beta\approx33, At\approx0.019, Ga\approx0.39$). The flow was initialized by imposing a $0.13D$-amplitude disturbance to the iso-surface $\hat{C}=0.5$. The `side-by-side' form of the disturbance, based on a product of sine and cosine functions, ensures that it vanishes at the wall, exhibits and antisymmetric distribution in a certain horizontal direction and a symmetric distribution in a second horizontal direction perpendicular to the first, thus mimicking the disturbance induced by the upside down flip of the experimental pipe. Given the large length-to-diameter aspect ratio of the pipe, $L/D\approx38$, simulating the entire pipe is very time consuming. To save computational resources, we decided to limit simulations to a time horizon at which the two fronts have separated by an approximate distance of $10D$. Preliminary runs were carried out with $L/D\approx30$ and the interface initially located $11D$ from the top of the pipe. Figure \ref{fig:axial} shows how the axial, i.e. vertical, velocity varies along the pipe axis at time $t/\tau=150$, when the two fronts are $10.5D$ apart. According to panel (a), the ascending and descending fronts have moved approximately $1.65D$ and $8.8D$, respectively. Moreover, panels (b-c) indicate that the velocity ahead of both fronts returns to zero within a $1D$ distance. Therefore, it can be concluded that the `active' part of the pipe within which the axial velocity is nonzero is approximately $12.5D$ long at $t/\tau=150$, with an `active' length less than $3D$ above the initial interface. Based on these results, the simulations described below were carried out in a pipe with $L/D=15$, with the initial interface being located $4D$ from the top end. \\
\indent To make the comparison of numerical predictions with experimental observations meaningful, it is essential to control the quality of the simulations, especially with respect to grid convergence issues. 
 To this end, we ran a first series of simulations with $512$ uniform cells distributed along the axial direction, corresponding to a vertical cell size $\Delta_V/D\approx0.0295$. Three different grids with $25\times25$, $52\times52$ and $79\times79$ uniform cells in the pipe cross section were employed, yielding dimensionless cell sizes in the horizontal plane $\Delta_{H}/D\approx0.046, 0.022$ and $0.0145$, respectively. \textcolor{black}{A fourth run was performed with the intermediate horizontal resolution and a twice as fine axial resolution, yielding a $52\times52\times1024$ grid with $\Delta_V/D\approx0.0146$. The resolution of that grid in each direction is virtually twice that of the coarse grid with $25\times25\times512$ cells.} \\
 \indent Computations were run on \textcolor{black}{the ADASTRA and OLYMPE clusters hosted by the CINES and CALMIP supercomputing centers, respectively}. One to three nodes and $54$ to $162$ cores were employed. \textcolor{black}{On the ADASTRA cluster}, the physical CPU time per time step and node for simulations using the $79\times79\times512$ grid with the anisotropic model was approximately $2.5$ seconds. The dimensionless time step $\Delta t/\tau$ was set to $1.56\times10^{-4}$, so that approximately $9.6\times10^5$ time steps were needed to reach a total time $t/\tau=150$, representing a total CPU time of $27$ days on $162$ cores.\\
  \begin{figure}
  \includegraphics[height=4.4cm]{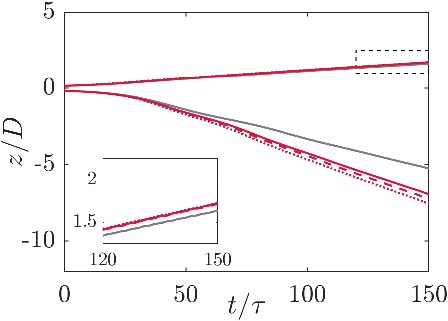}\hspace{-1mm}
        \includegraphics[height=4.4cm]{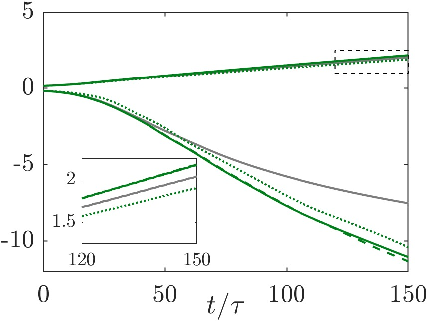}\hspace{-1mm}
         \includegraphics[height=4.4cm]{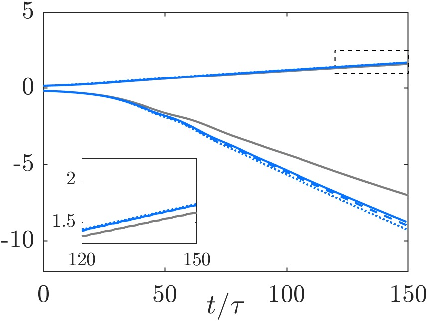}\\
        \vspace{2mm}
        \hspace{4mm}(a)\hspace{55mm}(b)\hspace{55mm}(c)
    \caption{Influence of grid resolution on the predicted position of the ascending and descending fronts ($\beta\approx33, At\approx0.019, Ga\approx0.39, L/D=15$). (a): arithmetic model $\tilde{\hat{\bm{\mathcal{T}}}}=2\lambda(\hat{C})\bm{\mathcal{S}}$; (b): harmonic model $\tilde{\hat{\bm{\mathcal{T}}}}=2\kappa(\hat{C})\bm{\mathcal{S}}$; (c) anisotropic model \eqref{Eq: Sigma_i}. In all panels, solid gray and red/green/blue lines refer to the $25\times25\times512$ and $52\times52\times512$ grids, respectively. The dashed and dotted lines refer to the $79\times79\times512$ and $52\times52\times1024$ grids, respectively. \textcolor{black}{The inset provides an enlarged view of the predictions for the ascending front in the late stage (dashed rectangle).} 
    }
    \label{fig:gridconv}
\end{figure}
\indent Figure \ref{fig:gridconv} displays the evolution of the ascending and descending fronts predicted by the three models on the four grids. Unsurprisingly, large variations are observed with each model between predictions obtained using the coarse and medium horizontal discretizations, with the speed of the descending front strongly increasing as the grid is refined. Some increase in the speed of the ascending front may also be discerned, especially with the harmonic model (Fig. \ref{fig:gridconv}(b)). With the harmonic and anisotropic models, only weak variations remain between predictions obtained using the $52\times52\times512$ and $79\times79\times512$ grids. In particular, the difference in the position of the descending front at $t/\tau=150$ is $1.5\%$ with the former model and $2.4\%$ with the latter one, while it is still $4.5\%$ with the arithmetic model. \textcolor{black}{Doubling the vertical resolution while keeping the medium $52\times52$ horizontal resolution is seen to have large effects in the case of the harmonic model, resulting in a marked decrease in the speed of both fronts. The situation is different with the arithmetic and anisotropic models. Indeed, no variation in the speed of the ascending front is noticed between the two vertical resolutions with both models. Conversely, doubling the vertical resolution increases the speed of the descending front by $8.7\%$ and $6.3\%$ in the case of the arithmetic and anisotropic models, respectively. \\
\indent The above observations lead to the following conclusions. Regarding the dynamics of the ascending front, grid convergence is reached with the arithmetic and anisotropic models on the $52\times52\times512$ grid. The same is not true regarding the descending front, since significant variations are still observed as the grid is refined, especially along the vertical direction. Assuming that variations resulting from changes in $\Delta_H$ or $\Delta_V$ may approximately be added, one can speculate that if simulations were run on a $79\times79\times1024$ grid, the speed of the descending front would be approximately $9\%$ larger than predicted on the $52\times52\times512$ grid with the anisotropic model, and $13.5\%$ larger with the arithmetic model. These estimates suggest that, although none of the models achieves a strict convergence on the finest grids considered, the anisotropic model is the closest to this objective. Moreover, comparing the three panels in Fig. \ref{fig:gridconv} reveals large differences in the final position reached by the descending front. Indeed, considering for instance results obtained on the $79\times79\times512$  grid and taking the prediction of the anisotropic model as reference, the vertical distance travelled by the front at $t/\tau=150$ is found to be $19\%$ shorter with the arithmetic model and $26\%$ larger with the harmonic model. If grid convergence were reached with all models, all three predictions would be identical. The large differences observed imply that at least two out of the three models are still far from convergence on the considered grid. The above observations suggest that this is the case with the arithmetic and harmonic \textit{ad hoc} models. Nevertheless, although predictions provided by the anisotropic model are closer to grid convergence, this is not sufficient to conclude that this model provides accurate predictions on any of the grids under consideration. This is where experimental data are required to play the role of `justice of the peace'.}\\

\end{document}